%% file: sigconf-sample.tex
\documentclass[sigconf]{acmart}
\usepackage{makecell}
\usepackage{multirow}
\usepackage{balance}
\usepackage{arydshln}
\usepackage{mdframed}

\AtBeginDocument{%
  \providecommand\BibTeX{{%
    \normalfont B\kern-0.5em{\scshape i\kern-0.25em b}\kern-0.8em\TeX}}}

\setcopyright{acmlicensed}
\copyrightyear{2018}
\acmYear{2018}
\acmDOI{XXXXXXX.XXXXXXX}

\acmConference[Conference acronym 'XX]{Make sure to enter the correct
  conference title from your rights confirmation emai}{June 03--05,
  2018}{Woodstock, NY}
\acmISBN{978-1-4503-XXXX-X/18/06}

\begin{document}

\title{Subtopic-aware View Sampling and Temporal Aggregation for Long-form Document Matching}


\author{Youchao Zhou, Heyan Huang, Zhijing Wu, Yuhang Liu, Xinglin Wang}
\affiliation{
 School of Computer Science and Technology, Beijing Institute of Technology \\ 
  \city{}
   \country{}
 }
\email{{yczhou,hhy63,zhijingwu,codelyh,wangxinglin}@bit.edu.cn}
\renewcommand{\shortauthors}{Zhou, et al.}

\begin{abstract}
Long-form document matching aims to judge the relevance between two documents and has been applied to various scenarios. Most existing works utilize hierarchical or long context models to process documents, which achieve coarse understanding but may ignore details. Some researchers construct a document view with similar sentences about aligned document subtopics to focus on detailed matching signals. However, a long document generally contains multiple subtopics. The matching signals are heterogeneous from multiple topics. Considering only the homologous aligned subtopics may not be representative enough and may cause biased modeling. In this paper, we introduce a new framework to model representative matching signals. First, we propose to capture various matching signals through subtopics of document pairs. Next, We construct multiple document views based on subtopics to cover heterogeneous and valuable details. However, existing “spatial aggregation” methods like attention, which integrate all these views simultaneously, are hard to integrate heterogeneous information. Instead, we propose “temporal aggregation”, which effectively integrates different views gradually as the training progresses. Experimental results show that our learning framework is effective on several document-matching tasks, including news duplication and legal case retrieval.

\end{abstract}

\ccsdesc[500]{Information systems~Retrieval models and ranking}

\keywords{ long-form document matching , clustering , subtopic analysis }



\maketitle


\input{sections/1_intro}
\input{sections/2_related}
\input{sections/3_method}
\input{sections/4_expset}

\input{sections/5_expres}
\input{sections/6_conclusion}


\bibliographystyle{ACM-Reference-Format}
\balance
\bibliography{sample-base}

\appendix
\input{sections/appendix}

\end{document}

%% file: sections/1_intro.tex
\section{Introduction}

\begin{figure}[t]
  \setlength{\belowcaptionskip}{-4pt} 
  \includegraphics[width=\linewidth]{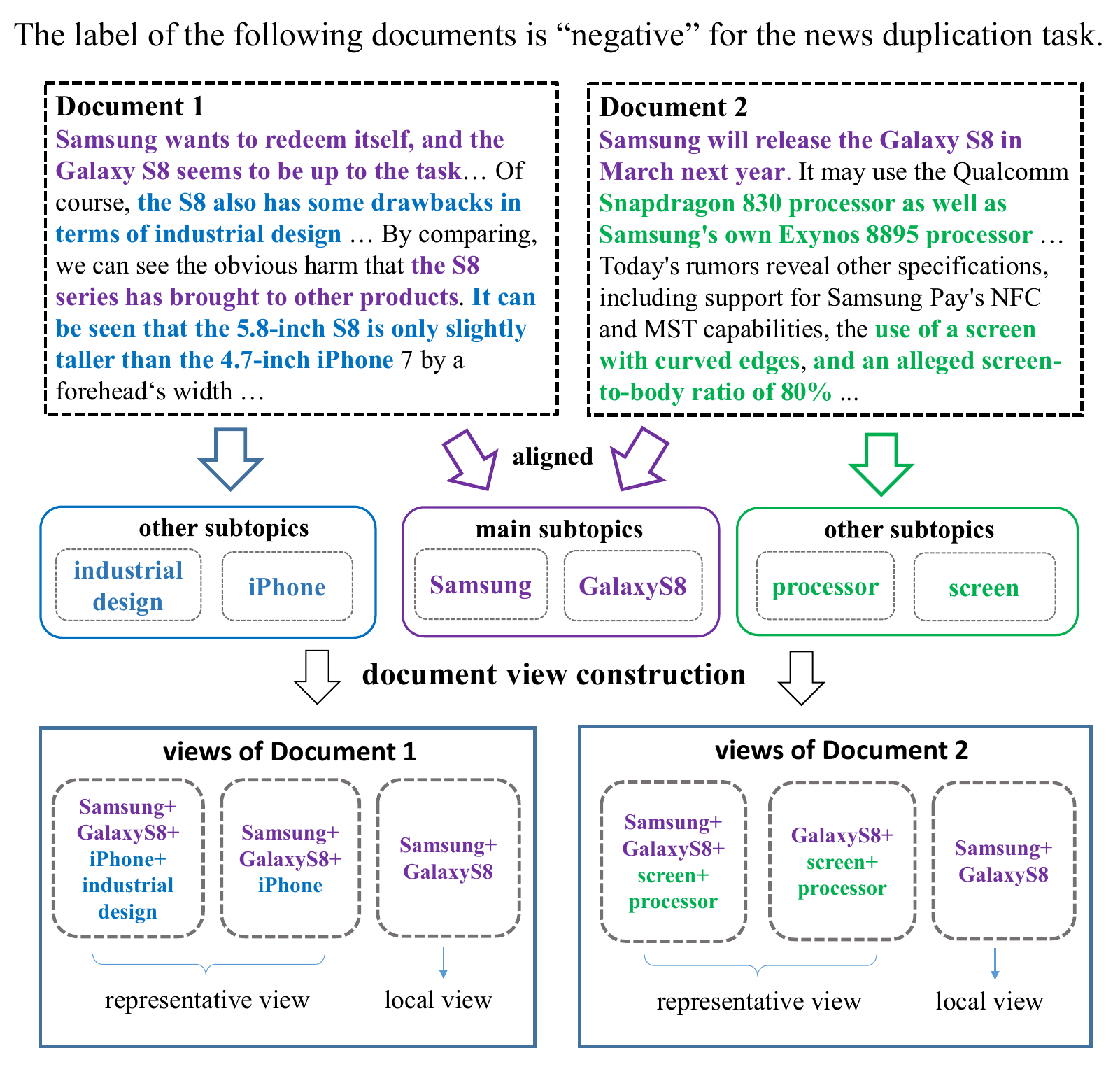}
  \caption{An example (translated from Chinese) of the news duplication task. The two documents describe different news events (the label is “negative”) but have aligned main subtopics. Existing methods pre-select sentences to form a single extracted summary (document view) and assume a local view that contains the most aligned part is salient and sufficient for the task. Learning such a homologous view with similar sentences may mislead the model. On the contrary, representative views combined with aligned subtopics and proper complementary subtopics indicate a mismatched relationship.}
  \label{fig: query_example}
\end{figure}

Long-form document matching has attracted increasing attention with the recent enhancement of the long-text processing ability of large language models. The task aims to judge whether two documents are relevant and has been applied for several scenarios like news duplication \cite{cig19acl,match_ig21cikm}, scientific document recommendation \cite{ncl22emnlp,scieval23emnlp}, and legal case retrieval \cite{parm22ecir,coliee23summary}.

Most existing methods contain a pre-trained model to encode documents. Since the document length generally exceeds the model's input limit, a straightforward solution is to use the truncated documents as input \cite{sdr21acl_finding}. However, truncation may cause the loss of valuable information \cite{psgrole19sigir,slr23tois}. Other solutions include using hierarchical models that split a document into smaller segments that models can process \cite{cda20emnlp,bertpli20ijcai,rprs24tois} or directly using long context models \cite{xlnet19nips,lawfm21}. These methods aim to increase the model's \textbf{receptive field} (i.e., the document content learned by the model) to cover more valuable information.

Furthermore, some researchers \cite{optmatch22coling,slr23tois} find that the irrelevant content in the document is redundant. As the model's receptive field increases, the noisy content may confuse the matching model \cite{match_ig21cikm}. Therefore, they \cite{match_ig21cikm,optmatch22coling} develop a content selection pipeline, which pre-selects salient sentences in documents to form extracted summaries and calculates a matching score based on these extracted summaries (we refer to a partial document, e.g. the extracted summary, as a document view). These methods assume that salient parts with aligned sentences (similar in a general sense) can reflect the main topics of the document and are enough to model the matching signal. However, a long document generally contains multiple main topics and subtopics \cite{texttilling,sector19tacl}. The matching signals are heterogeneous from multiple topics. Considering only the homologous main topics may not be representative enough and may cause biased modeling.

Taking a document pair from the news duplication task in Figure \ref{fig: query_example} as an example. Document 1 primarily introduces the business situation of Samsung Company, and Document 2 presents the premature disclosure of the “Galaxy S8”. They have the same main subtopics, “Samsung, GalaxyS8”, but describe different news events; therefore, they are labeled with “negative” in the CNSE dataset \cite{cig19acl}. Existing methods tend to form two local views with sentences about the aligned main subtopic and ignore the essential background knowledge, leading to biased modeling. On the contrary, representative views, combining main and essential complementary subtopics with fewer similarities, are more comprehensive for the task.

\begin{figure*}[t]
  \vspace{-1em}
  \includegraphics[width=\linewidth]{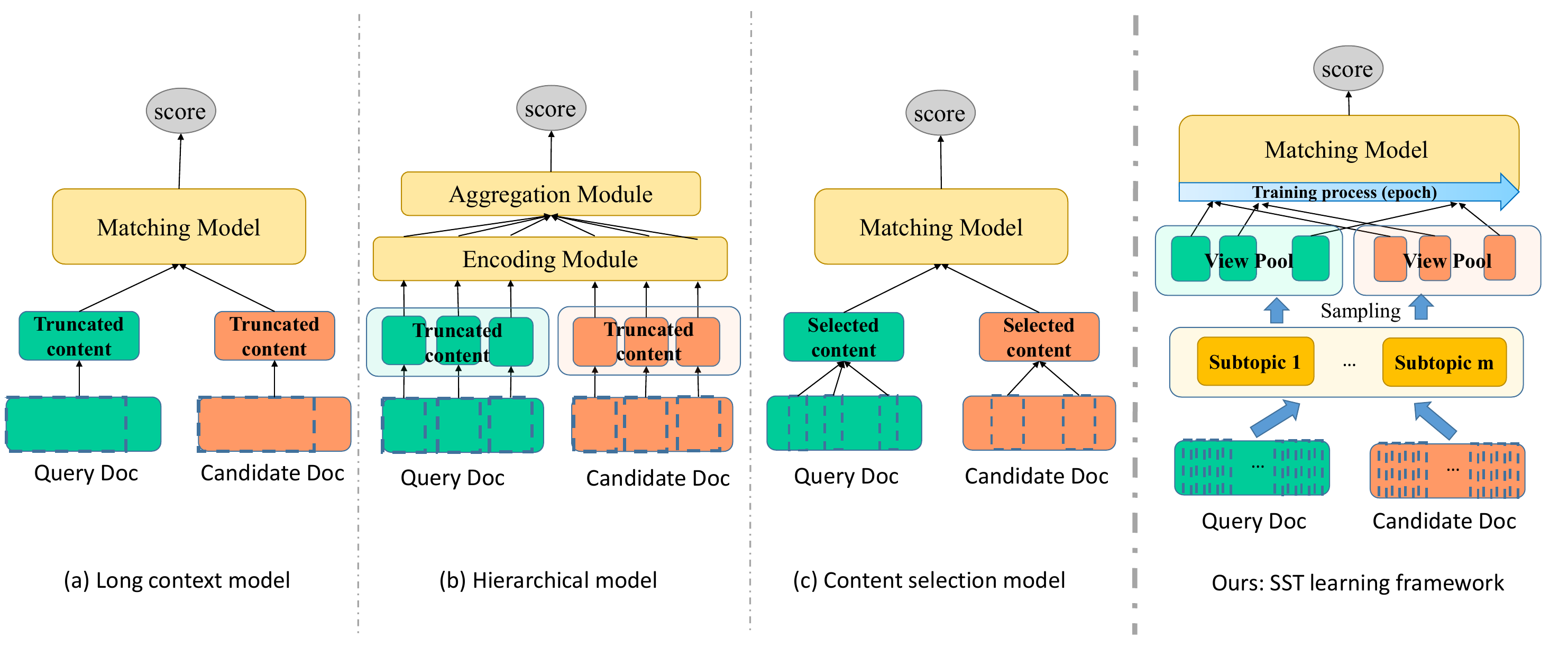}
  \caption{The comparison of different model architectures. }
  \label{fig: method comparison}
\end{figure*}

Therefore, to achieve comprehensive and detailed modeling, we propose a new learning framework named SST, including a \textbf{S}ubtopic-aware view \textbf{S}ampling strategy and \textbf{T}emporal aggregation strategy. The first strategy includes the subtopic discovery step and the view sampling step. The subtopic discovery step clusters fine-grained linguistic units (use sentences here) to analyze the subtopic modularity \cite{dmon23jmlr}, including alignment and complement relationship between documents. Meanwhile, we propose an adaptive clustering method to better capture high-order sentence relationships. Based on different assumptions about representativeness, we design various sampling methods to construct views, which include comprehensive valuable matching signals.

Moreover, as these views contain various signals and vary in representativeness subtly, synthesizing them remains challenging. Recent research \cite{exdocrank23tois} validates that even valuable signals remain redundant for models. The finding indicates that the model is hard to distinguish the details of these views. Therefore, “spatial aggregation” methods like attention \cite{smash19www} are inappropriate because they integrate all these views simultaneously. Instead, our second strategy gradually integrates different views as the training progresses. It is inspired by Dropout \cite{dropout}, which disables neurons in a network layer randomly. The Dropout process amounts to sampling a sub-network from a larger network and could be regarded as an ensemble learning process. Specifically, we train a view once but with different views of a document pair, which effectively ensemble the subtle valuable details.

We validate our learning framework on two domains of long-form document tasks, i.e., the news domain and the legal domain. To summarize, our contributions are as follows:

\begin{itemize}
\item We introduce a new learning framework SST, (\textbf{S}ubtopic-aware view \textbf{S}ampling and \textbf{T}emporal aggregation), including a strategy that build representative view and an aggregation strategy enabling the effectively model learning.

\item We propose variants of the subtopic discovery step and the view sampling step to capture heterogeneous valuable matching signals. These provide insights into selecting valuable information for long-form document matching tasks.

\item Experimental results show that our learning framework is effective on multiple long-form document matching tasks. We conduct further analysis to reveal how the two strategies of the SST framework collaborate. \footnote{ The code is available at https://anonymous.4open.science/r/SST-learning-framework-C4C2/}
\end{itemize}

%% file: sections/2_related.tex
\section{Related work}

\subsection{Document Matching}

Document matching, including ranking and classification, has been extensively researched by many researchers. Most of these studies \cite{pcfg20www,Intra-select21sigir} focus on ad-hoc document ranking tasks, which aim to match short queries to long documents.  However, long-form document matching is more complicated as there exist multiple subtopics in documents for matching models to distinguish. Researchers mainly leverage three streams of models, namely hierarchical models, long-context models, and content selection models. The first stream \cite{smash19www,smith20cikm,cda20emnlp,parm22ecir} leverage hierarchical models, which split the document into multiple segments to get representation and aggregate them to get a matching score.  The second stream  uses the straight-forward long-context models which directly extend the context window by variant attention mechanisms \cite{xlnet19nips,lawfm21,fot24nips}. However, they are not as effective as expected in long-form documents as the limited document-level supervision still hinders learning the matching signal. 
Compared to these methods, the content selection models \cite{match_ig21cikm,optmatch22coling,slr23tois}, which select salient content in documents to form a short document view, have shown better performance. For example, \citet{match_ig21cikm} apply PageRank algorithm on both sentence and word level to avoid noise and \citet{optmatch22coling} leverage optimal transport theory to filter out non-salient sentences. 
Recently, researchers adopt powerful LLMs for ranking tasks \cite{rankgpt23emnlp} which also has lengthy context window. \citet{rankllama23sigir} directly finetune the LLaMa model in rankT5 style \cite{rankt5_23sigir} and get better results than encoder-only models. Some works \cite{rlcf24sigir,legalsum23sigir_ap} use LLMs to analysis document and condense the salient part for further ranking. We also test LLM-based model in our experiments.

\subsection{Document Subtopic Analysis}

Document naturally contains structure and topic transition as it expands \cite{texttilling}. As an essential subject, document analysis has been investigated by different research communities \cite{s2s23kdd,segformer23aaai,topiccohere_23emnlp}. To avoid conceptual confusion, we refer to the continuous semantic fragments as “subtopic” according to \cite{textrank}. The mainstream methods can be categorized into supervised or unsupervised. The supervised methods \cite{s2s23kdd,segformer23aaai} generally work on more structured Wikipedia, which performs better than unsupervised ones. However, they \cite{sector19tacl} construct labels with complicated pipelines and cannot easily transfer to arbitrary domains. The unsupervised methods \cite{graphseg16acl,topictiling,texttilling} have more application scenarios. It separates the more similar and continuous sentences into the same subtopic and thus makes different subtopics more dissimilar. 

However, the specific form of the subtopic varies, such as multi-document summarization \cite{subtopic_mdsum_19_emnlp}, search result diversification \cite{subtopic_retrieval_15,multi_grain_srd24tois}. For example, in multi-document summarization, \cite{subtopic_mdsum_19_emnlp} considered the subtopic of each sentence as a hidden variable and implicitly represented as a subtopic vector. For search result diversification, the subtopic of the document is explicitly defined as the Google suggestions of queries \cite{multi_grain_srd24tois}. In our task, the key idea is to find valuable matching signals and decide the document-level relationship. Finding modularity information, including alignment and complement between document pairs is essential. Meanwhile, the key idea is quite similar to clustering, which analyzes the modularity of data samples \cite{multidoc_cluslink08sigir,dmon23jmlr}. The difference is that a subtopic has continuous sentences in the document, and clustering has no continuity limitation. Therefore, we leverage clustering methods to analyze the subtopic modularity between document pairs. In our tasks, we denote the cluster of document pairs as the subtopic.

%% file: sections/3_method.tex
\section{Methodology}

\begin{figure*}[t]
  \vspace{-1em}
  \includegraphics[width=\linewidth]{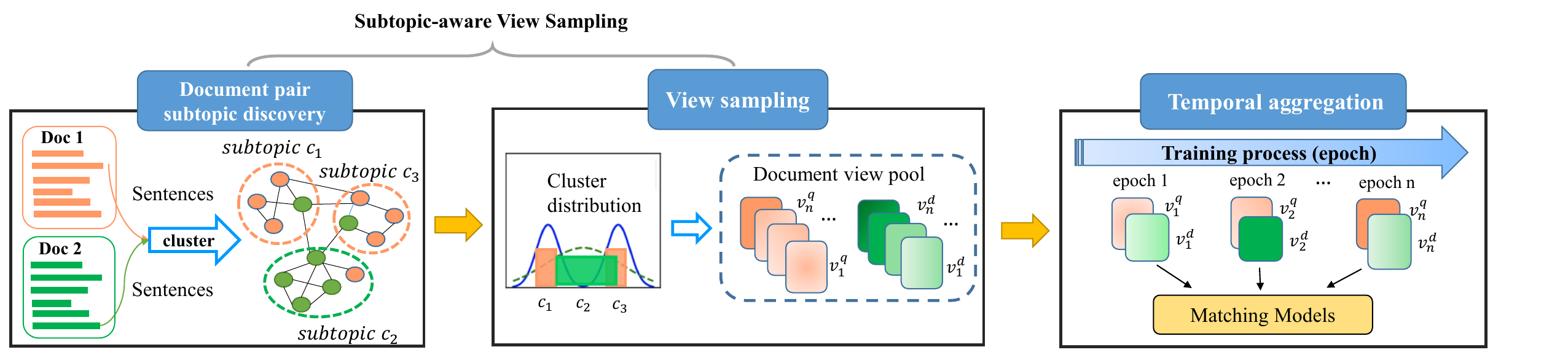}
  \caption{The process of our SST framework. It contains two strategies: subtopic-aware view sampling and temporal aggregation. The first strategy contains two steps: the document pair subtopic discovery and the view sampling. }
  \label{fig: method overview}
\end{figure*}

\subsection{Preliminaries}
Document matching tasks include classification and ranking tasks following a similar task format. Given a query document $q$ and a candidate document set containing $n$ documents $\{ d_1,d_2,...,d_p \}$, the matching model $Match(q,d)$ needs to measure the similarity between $q$ and a candidate document $d$. The matching model can be a cross-encoder or bi-encoder. The evaluation metrics include classification-based metrics when $p=1$ and ranking-based metrics when $p>1$. As a document in the long-form document matching task typically exceeds the model input size, the input query and document are generally further processed. The methods can be categorized into three main streams as shown in Figure \ref{fig: method comparison}. The first leverages long-context models to process documents. The second splits a document into smaller sequential units that short-context models can process. Then, they use aggregation modules to get final scores. These two methods achieve a coarse understanding of the document. The third leverages content selection methods to select the salient document parts for the short-context model, which focuses more on valuable local details. Unlike existing methods, our framework captures comprehensive and detailed matching signals between documents. The framework is illustrated in Figure \ref{fig: method overview}. It consists of two strategies, and the first one contains two steps. The details are described as follows.  

\subsection{Document Pair Subtopic Discovery}
\label{sec:acs}

To capture subtopics modularity, including sentence alignment and supplement between documents, we leverage clustering methods. We treat clusters as “subtopics” because a document subtopic consists of semantic coherence sentences \cite{texttilling} and has equivalent meaning with clusters. We denote the query as $q=\{s_1^q,s_2^q,...,s_{l_q}^q \}$ and candidate document $d=\{s_1^d,s_2^d,...,s_{l_d}^d \}$, where $l_q$ and $l_d$ are the size of sentence set. The cluster process happens on the combined sentence set $S=\{ s_1^q,s_2^q,...,s_{l_q}^q,$ $s_1^d,s_2^d,...,s_{l_d}^d \}$. The sentences of two docs in the same cluster reflect alignment, and the different clusters for the same doc reflect supplement. Specifically, we give two variants: an efficient direct implementation that leverages off-the-shelf clustering methods and an adaptive clustering implementation that better captures the high-order sentence relationship between documents. They are described as follows.

\textbf{Direct Clustering} 

Previous studies \cite{optmatch22coling,match_ig21cikm} suggest that the more similar parts between documents are more informative. The direct implementation follows the idea.
As clustering methods rely on the similarity measure among the sentences, we first build sentence similarity matrix $A$ as in TextRank \cite{textrank}:

\begin{equation}
A_{ij}\!=\!\frac{| \{ w| w\! \in s_i, w\! \in s_j\! \}|} {  log(|s_i|)\!+\!log(|s_j|)\! }, s_i,s_j \in  S,
\end{equation}
where $A_{ij}$ denotes the similarity between $s_i$ and $s_j$, $w$ denotes a word in sentences, and $|\cdot|$ denotes the size of the word set or length of sentences. Then, we can efficiently leverage the similarity $A$ and spectral clustering algorithm \cite{spectral} to get the clusters set $C$ as follows:
\begin{equation}
    C = Spectral(A) =\{c_1,c_2,..,c_m \},
\end{equation}
where $m$ is the number of clusters. Compared to other clustering algorithms like k-means or hierarchical clustering that suit a large sample size of more than thousands of samples, we choose spectral clustering \cite{spectral} as it suits the scale of our problem with tens of sentences.



\textbf{Adaptive Clustering} 

Though the direct implementation uses a general similarity measure, it may not be suitable for document matching. In a specific context, the meaning of a sentence can change, especially under the domain-specific document matching task. As Figure \ref{fig: query_example} shows, seemingly similar sentences have negative meanings when considering the high-order sentence relationship (supplement subtopics). To provide a better measurement among sentences, we propose to optimize the sentence representation with designed adaptive clustering loss.

First, we encode the combined sentences set $S$ with a pre-trained language model into embedding $Emb=\{ e_1^q,e_2^q,...,e_{l_q}^q,e_1^d,e_2^d,...,e_{l_d}^d \}$. Then we build sentence similarity matrix $A$ with commonly used dot product:
\setlength{\jot}{-2pt}
\begin{equation}
A_{ij}\!=e_i \cdot e_j , e_i,e_j \in  Emb,
\end{equation}
where $A_{ij}$ denotes the similarity between $e_i$ and $e_j$.
Next, we train an unsupervised clustering metrics model $B=Cluster_u(A)$ on document matching datasets, where $B \in \mathcal{R}^{((l_q+l_d) \times m)}$ indicate the sentences allocation to $m$ clusters. Unlike traditional clustering methods, neural clustering models enable the whole optimization process and show comparable performance \cite{unspecral20icml,sdcn20www}.

As the sentence meanings vary with the document relationship, we designed the adaptive clustering loss for positive and negative document pairs correspondingly.
For positive pairs, dissimilar supplementary sentences may also have positive meanings. To reallocate these potentially aligned sentences, we design the loss $\mathcal{L}_p$:
\begin{gather}
    P^{cq} = \frac{\sum_{i=1}^{l^q} Max(B_i)}{l^q} , P^{cd} = \frac{\sum_{i=l_q+1}^{l^q+l^d} Max(B_i)}{l^d} \notag \\
    \mathcal{L}_p (P^{cq} , P^{cd} ) = JS(P^{cq} || P^{cd})
    = \sum_{i=1}^{m} \frac{1}{2} (p_{i}^{cq} log \frac{p_{i}^{cq}}{p_{i}^{cd}} + p_{i}^{cd} log \frac{p_{i}^{cd}}{p_{i}^{cq}} ),   \notag \\
\end{gather}
where the $B_i$ is the probability allocation of the sentence $i$ to $m$ clusters, $Max$ is max-poolng operation and $JS$ is the Jensen-Shannon divergence. $\mathcal{L}_p$ assume that the two documents have a similar structure, which regularizes supplement subtopics through the coarse matching relationship.

For negative pairs, similar sentences also have negative meanings. However, we do not need to distinguish them thoroughly here. To alleviate the burden of the matching model, we expect those hard-to-distinguish sentences to be in aligned subtopics while the easy ones are the opposite. The loss $\mathcal{L}_n$ is:
\begin{gather}
    \mathcal{L}_n= || B^TB-I||_F,
\end{gather}
where $|| \cdot ||$  is the Frobenius norm, which encourages the cluster distribution matrix to be orthogonal \cite{unspecral20icml} and form more distinguished clusters.

Overall, the clustering head of $Cluster_u(A)$ is frozen once it pre-trained and the training loss $\mathcal{L}_u$ is as follows:
\begin{gather}
    \mathcal{L}_u=\mathcal{L}_p + \lambda \mathcal{L}_n ,
\end{gather}
where $\lambda$ is a hyper-parameter.

\subsection{View Sampling}
The subtopic discovery step provides modularity information between documents, which enables us to build representative views for a fine-grained learning process. We design three sampling methods to obtain document views based on different assumptions of representativeness. They are uniform, hard, and soft sampling. 

The first sampling method is \textbf{uniform sampling}, which assumes each subtopic is equally essential. This form a heterogeneous view containing various details. We sample sentences from each subtopic follow a uniform distribution:
\begin{gather}
    P^u_{i} = \frac{1}{m} (1 \leq i \leq m), \notag \\
    v^q=rand(C^q,k,P^u),v^d=rand(C^d,k,P^u),
\end{gather}
where $v^q$/$v^d$ is the view of the original document $q$/$d$. The operation $rand(C,k,P)$ means choosing $k$ clusters (allow for repetition) using the roulette wheel selection method  \cite{Roulette-wheel} following the distribution $P$, where we further randomly sample one sentence from each selected cluster. The sampled $k$ sentences form a document view $v$.

The second sampling method is \textbf {hard sampling}, which assumes the primary aligned subtopics contain representative details. As a larger cluster naturally contains more information, the largest subtopics can be regarded as the primary subtopics. This sampling method focuses on the primary subtopics, which form a homologous view containing similar details. It can be represented as:
\begin{gather}
    M=\underset{1 \leq i \leq m} {argmax} (min(|c_i^q|,|c_i^d|)), \notag \\
    v^q=rand(\{c_M^q\},k,1),v^d=rand(\{c_M^d\},k,1),
\end{gather}
where $m$ is the cluster size. Notice that document lengths and structure vary, so we can not ensure that the largest cluster satisfies the view size. In this case, we use uniform sampling in other clusters until we fulfill a document view.

The third sampling method is \textbf{soft sampling}, which additionally assumes the aligned and supplementary subtopics all contain valuable details. The sampling method form views of moderate heterogeneous details. It is as follows:
\begin{gather}
    P^{q}_{i} = \frac{|c^q_i|}{l_q},
    P^{d}_{i} = \frac{|c^d_i|}{l_d},
    P^{a}_{i}= \frac{min(|c_i^q|,|c_i^d|)}{\sum_{j=1}^{m} min(|c_j^q|,|c_j^d|)}, \notag \\
    Q^{q} = \frac{P^{a} * P^{q} }{P^a \cdot P^{q}}, 
    Q^{d} =\frac{P^{a} * P^{d} }{P^a \cdot P^{d}}, \notag \\
    v^q=rand(C^q,k,Q^q),v^d=rand(C^d,k,Q^d),
\end{gather}
where $*$ is element-wise product and $\cdot$ is dot product. The $Q$ distribution considers both the importance of aligned subtopic between documents ($P^a$) and the subtopic of each document ($P^q/P^d$).

\subsection{Temporal Aggregation}

The document views are potentially representative of the task. They contain subtly various detailed information and are consequently hard to aggregate. To synthesize them effectively, we propose the \textbf{temporal training strategy}. It differs from traditional “spatial aggregation” methods like hierarchical models, which leverage pooling or attention methods to aggregate multiple segment information. As for temporal aggregation, we progressively ensemble the multiple document views in the temporal domain. Specifically, we leverage above-desinged view sampling methods to get a view pool $V(q,d)\!=\!\{(v^q_1,v^d_1),...,(v^q_E,v^d_E)\}$, where $E$ is the max training epoch. In the $t^{th}$ training epoch, the matching loss $\mathcal{L}_m$ for document pair $(q,d)$ as follows:
\begin{equation}
\mathcal {L}_m^t
= \mathcal{L}(s(v^q_t,v^d_t),l),
\end{equation}
where $\mathcal{L}$ is the loss for classification (e.g., Cross Entropy) or ranking (e.g., InfoNCE) correspondingly,  $l$ is the corresponding label, and $v^q_t,v^d_t$ is the $t^{th}$ view of $V(q,d)$. The strategy is model-independent and thus has a wider range of applications. 

As the model trains with multiple views, which covers a wide range of a document and thus gets a more comprehensive understanding of document pairs, we also build a view pool size $n$ in the inference stage. For classification tasks, the aggregation prediction $s_c$ is:
\begin{equation}
s_c=Max(s(v^q_1,v^d_1),...,s(v^q_n,v^d_n)),
\end{equation}
where $Max$ is max-pooling. We use max-pooling as it is insensitive to noise, and eliminating distractive views is more important for classification. For ranking tasks, the aggregation prediction is:
\begin{equation}
s_r=Mean(s(v^q_1,v^d_1),...,s(v^q_n,v^d_n)),
\end{equation}
where $Mean$ is mean-pooling. We use mean-pooling as views all provide information and the mean-pooling provide a fine-grained score for ranking tasks. 

%% file: sections/4_expset.tex
\section{Experiment Settings}

\subsection{Datasets}
\label{sec:appdenix_dataset}
We conduct experiments on news duplication and legal case retrieval. The statistics of the four datasets are shown in Table \ref{tab:stat}.
For news duplication, we use the CNSE and CNSS datasets \cite{cig19acl}. The task of CNSE is to identify whether a pair of news articles report the same breaking news (or event), and the CNSS is to identify whether they belong to the same series of news, which is a broader concept. Notice the negative documents in the two datasets are not randomly generated. Document pairs that contain similar keywords are selected, which brings challenges. 

Legal case retrieval aims to retrieve relevant cases to a given query case, whose definition of relevance is more complex and domain-specific than generic ad hoc retrieval tasks. The relevance is more related to the constitutional element that describes the fact of the criminals \cite{sailer23sigir}. For this task, we conduct an evaluation on two datasets: the LeCaRDv2 dataset \cite{lecardv2_24sigir} and the COLIEE 2023 dataset \cite{coliee23summary}. 

\begin{table}[t]
  \centering
  \caption{Statistics of the four datasets.}
    \scalebox{0.6}{
    \begin{tabular}{p{13.125em}|cccccccc}
    \toprule
    \multicolumn{1}{c}{Datasets} & \multicolumn{2}{c}{CNSE} & \multicolumn{2}{c}{CNSS} & \multicolumn{2}{c}{LeCaRDv2} & \multicolumn{2}{c}{COLIEE2023} \\
    \multicolumn{1}{c}{Data splits} & train & test  & train & test  & train & test  & train & test \\
    \midrule
    \midrule
    \# Query docs & 17438 & 5812  & 20102 & 6700  & 640   & 160   & 959   & 319 \\
    \# Candidate docs per query & 1     & 1     & 1     & 1     & 55192 & 55192 & 4400  & 1335 \\
    \# Relevant doc & 7762  & 2548  & 10039 & 3407  & -     & -     & -     & - \\
    \# Avg. relevant doc per query & -     & -     & -     & -     & 23.78 & 24.35 & 4.68  & 2.69 \\
    \# Avg. sentences per doc & 20.1  & 20.83 & 20.4  & 21.82 & 70.63 & - & 42.23 & 44.33 \\
    \# Avg. sentence length ( \# word) & 52.7  & 55.78 & 57.27 & 56.6  & 67.5  & - & 21.51 & 29.37 \\
    \bottomrule
    \end{tabular}%
    }
  \label{tab:stat}%
\end{table}%

\subsection{Baseline Models}
\label{sec:appendix_baseline}
We compare our model with traditional models and neural document matching models. They are described as follows.
\begin{itemize}
\item Traditional lexical models: BM25 \cite{bm25}, BERTTopic \cite{berttopic}.
\item Truncation models: BERT \cite{bert}, CoLDE \cite{colde23tkdd}, SAILER \cite{sailer23sigir}.
\item Long-context models and LLM-based models: XLNeT \cite{xlnet19nips}, Lawformer \cite{lawfm21}, Baichuan-7B \cite{baichuan23}, GPT-3.5-turbo-0125 (4K), GPT-4-0125(4K).
\item Hierachical models: BERT-PLI \cite{bertpli20ijcai}, RPRS \cite{rprs24tois}.
\item Content selection models: CIG \cite{cig19acl}, Match-Ignition \cite{match_ig21cikm}, OPT-Match \cite{optmatch22coling}.
\end{itemize}
As the CIG, Lawformer and SAILER are designed for specific domains, we report their results on their corresponding tasks. The details are in the Appendix \ref{App:baseline models}.

\subsection{Implementation Details}
\label{sec:appendix_implement}
We report the default settings for our framework. For the direct clustering methods, we use sklearn tools \footnote{ https://scikit-learn.org/stable/modules/generated/sklearn.cluster.SpectralClustering.html}. For the adaptive clustering methods, we consider the DMoN \cite{dmon23jmlr} models as unsupervised metrics and set $\lambda$ in \{0.1,1,10\} for training. We set the receptive field of models to 40 sentences (the maximum sentence number for a document). Instead of setting the same clustering number for each document pair, we set it according to an expected average cluster size. The default is 6 sentences for each cluster. For aggregation inference, we set the view pool size as 3. The classification metrics are reported with sklearn tools \footnote{ https://scikit-learn.org/stable/api/sklearn.metrics.html}, and the ranking metrics are reported with trec-eval tools \footnote{ https://github.com/usnistgov/trec\_eval }. More details are in Appendix \ref{App:implementation details}.

\begin{table}[t]
  \centering
    \caption{Main results on news duplication task. The results of CIG and Match-Ignition are from the original paper. For LLM-based models, $z$ denotes zero-shot evaluation, and $f$ denotes fine-tuning results. The best results are marked in bold, and the best of non-SST methods are underlined. For the SST framework,$d/a$ denotes direct/adaptive clustering, and $h/s$ denotes hard/soft sampling. Compared to all other replication results, the top-3 SST results are statistically significant under pair-wise t-test (p-value < 0.05). }
    \begin{tabular}{cp{6.5em}llll}
    \toprule
    \multicolumn{2}{c}{\multirow{2}[4]{*}{Models}} & \multicolumn{2}{c}{CNSE} & \multicolumn{2}{c}{CNSS} \\
\cmidrule{3-6}    \multicolumn{2}{c}{} & Acc   & F-1   & Acc   & F-1 \\
    \midrule
    \midrule
    \multirow{2}[2]{*}{I} & \multicolumn{1}{l}{BM25} & 69.63 & 66.62 & 67.66 & 70.41 \\
          & \multicolumn{1}{l}{BERTopic} & 67.43 & 65.85 & 69.28 & 72.45 \\
    \midrule
    \multirow{3}[2]{*}{II} & \multicolumn{1}{l}{BERT} & 83.97 & 81.67 & 89.63 & 90.09 \\
          & \multicolumn{1}{l}{XLNeT} & 83.21 & 81.09 & 90.22 & 90.28 \\
          & \multicolumn{1}{l}{CoLDE} & 84.47 & 82.18 & 89.97 & 90.42 \\
    \midrule
    III   & \multicolumn{1}{l}{BERT-PLI} & 77.90 & 75.73 & 85.95 & 86.22 \\
    \midrule
    \multicolumn{1}{c}{\multirow{4}[2]{*}{IV}} & \multicolumn{1}{l}{CIG} & 84.64 & 82.75 & 89.77 & 90.07 \\
          & \multicolumn{1}{l}{CIG-$Sim_g$} & 84.21 & 82.46 & 90.03 & 90.29 \\
          & \multicolumn{1}{l}{OPT-Match} & 83.80 & 81.88 & 90.81 & 90.97 \\
          & \multicolumn{1}{l}{Match-Ignition} & \underline{86.32} & \underline{84.55} & 91.28 & 91.39 \\
    \midrule
    \multirow{4}[2]{*}{V} & Baichuan-$7B^z$ & 43.75 & 60.85 & 50.82 & 67.37 \\
          & \multicolumn{1}{l}{GPT-$3.5^z$} & 66.67 & 59.29 & 62.66 & 70.94 \\
          & \multicolumn{1}{l}{GPT-$4^z$} & 77.14 & 77.62 & 80.12 & 80.72 \\
          & Baichuan-$7B^f$ & 84.67 & 82.56 & \underline{91.82} & \underline{92.22} \\
    \midrule
    \midrule
    \multirow{4}[2]{*}{VI} & $BERT + SST^{dh}$ & 86.25 & 85.10 & 92.52 & 92.71 \\
          & \multicolumn{1}{l}{ $BERT + SST^{ds}$ } & 86.91 & 85.69 & 92.76 & 92.90 \\
          & \multicolumn{1}{l}{$BERT + SST^{ah}$} & 87.04 & 85.84 & 93.01 & 93.24 \\
          & $BERT + SST^{as}$ & \textbf{87.65} & \textbf{86.45} & \textbf{93.45} & \textbf{93.58} \\
    \bottomrule
    \end{tabular}%
  \label{tab:news_results}%
\end{table}%

%% file: sections/5_expres.tex
\section{Experimental Results}
In this section, we first compare our framework with other mainstream models. Then, we conduct an ablation study and analyze our framework with different hyper-parameters. Finally, we give further analysis and conduct a case study to show how our framework 
synthesis representative views to achieve comprehensive and detailed model learning.

\subsection{Main Results}
We present our methods based on hard and soft sampling here as we find they perform better. The results of uniform sampling are shown in section \ref{sec:abla}.

The main results on the news duplication task are present in Table \ref{tab:news_results}. There are mainly five main categories of baseline models. Category (I) is the traditional model, which leverages word overlap or topic model to get document representation and leverages classification model to get results. They do not show good results, which means the coarse representation provides limited information. Category (II) is the traditional pre-trained model. Though XLNeT has an infinite context window and BERT has a limited context window, the latter performs better. This indicates that the long-context model can not effectively distinguish valuable information from long documents. Category (III) is the hierarchical model with a longer receptive field than BERT. BERT-PLI does not perform well on this task. It probably suits the task where similar coarse information reflects the document-level relationship as legal case retrieval. Category (IV) is the content selection model, and CIG is the new-specific method that pre-selects salient keywords to boost the results. Though these methods have a similar context window as BERT, they perform better as they focus on the potentially more salient part. The category (V) is the prevalent LLM-based model, with powerful understanding ability and much longer context windows. The Baichuan-7B model performs well only under the supervised finetune settings. Though it performs better compared to BERT and XLNET, the model parameter quantity is far greater than that of the small language model. However, the zero-shot performances of the open-source or powerful GPTs are not well. The domain characteristics may contribute to the results as the category (I) model exhibits. Compared to content selection methods, our method shows generally better performance. 

Table \ref{tab:legal_results} presents the main results on legal case retrieval. The two-stage experiment is based on the top 40 for LeCaRDv2 and the top 20 for COLIEE 2023 BM25 ranking cases as the first-stage ranking result. The general trend is different from news tasks. The content selection methods (category III) implemented based on the legal pre-trained model SAILER perform worse. Generally, our methods improve based on SAILER, which shows an orthogonal growth in the re-ranking stage. We give a discussion in section \ref{sec:abla}. 

In both tasks, soft sampling generally performs better than hard sampling, indicating the importance of moderate heterogeneous and aligned salient content. We give more detailed results on various document lengths as in Figure \ref{fig: details result}. We observe that with our SST framework, the trained model performs better on long-context documents and even on short-context documents. It indicates that our framework enhances the model's comprehensive and detailed understanding of documents.

\begin{table}[t]
  \centering
  \setlength{\belowcaptionskip}{-2pt}
  \caption{Main results on legal case retrieval. The best results are marked in bold, and the best of non-SST methods are underlined. The top SST results are statistically significant under pair-wise t-test (p-value < 0.05) compared to all other replication results.}
  \scalebox{0.8}{
    \begin{tabular}{ll|cccc}
    \toprule
    \multicolumn{2}{c|}{\multirow{2}[4]{*}{Models}} & \multicolumn{2}{c}{LeCaRDv2} & \multicolumn{2}{c}{COLIEE 2023} \\
\cmidrule{3-6}    \multicolumn{2}{c|}{} & NDCG@10 & NDCG@20 & NDCG@5 & NDCG@10 \\
    \midrule
    \midrule
    \multicolumn{2}{c|}{\textbf{One-stage}} &       &       &       &  \\
    \multicolumn{1}{c}{\multirow{2}[1]{*}{I}} & BM25  & 47.80 & 42.08 & 30.02 & 34.25 \\
          & SAILER & 49.29 & 44.62 & 29.87 & 32.31 \\
    \midrule
    \midrule
    \multicolumn{2}{c|}{\textbf{Two-stage}} &       &       &       &  \\
    \multirow{4}[1]{*}{II} & Lawformer & 47.74 & 39.68 & -     & - \\
          & BERT  & 45.48 & 38.15 & 25.77 & 27.46 \\
          & SAILER & \underline{50.87} & \underline{47.44} & \underline{32.23} & \underline{36.51} \\
          & CoLDE & 48.14 & 45.06 & 31.56 & 35.76 \\
    \midrule
    \multirow{2}[2]{*}{III} & Match-Ignition & 48.35 & 44.92 & 28.67 & 33.23 \\
          & OPT-Match & 49.49 & 46.24 & 30.36 & 34.55 \\
    \midrule
    \multicolumn{1}{c}{\multirow{2}[2]{*}{IV}} & SAILER-PLI & 48.56 & 44.90 & 29.87 & 33.61 \\
          & RPRS  & 49.73 & 44.71 & 31.60 & 36.69 \\
    \midrule
    \midrule
    \multirow{4}[2]{*}{V} & SAILER + $SST^{dh}$ & 50.46 & 47.19 & 32.05 & 36.34 \\
          & SAILER + $SST^{ds}$ & 51.82 & 49.25 & 33.09 & 37.96 \\
          & SAILER + $SST^{ah}$ & 51.37 & 48.73 & 32.68 & 37.02 \\
          & SAILER + $SST^{as}$ & \textbf{53.19} & \textbf{50.36} & \textbf{33.43} & \textbf{38.57} \\
    \bottomrule
    \end{tabular}%
    }
  \label{tab:legal_results}%
\end{table}%

\begin{table}[t]
  \setlength{\belowcaptionskip}{-3pt} 
  \centering
  \caption{Ablation study and variant study of modules.}
  \scalebox{0.9}{
    \begin{tabular}{lcccc}
    \toprule
    \multicolumn{1}{c}{\multirow{2}[2]{*}{Models}} & \multicolumn{2}{c}{CNSE} & \multicolumn{2}{c}{COLIEE 2023} \\
          & Acc   & F-1   & NDCG@5 & NDCG@10 \\
    \midrule
    BERT/SAILER + $SST^{as}$ & 87.65 & 86.45 & 33.63 & 38.57 \\
    \midrule
    \midrule
    w/o aggregation inference & 87.28 & 86.01 & 33.07 & 38.02 \\
    w/o temporal aggregation & 85.37 & 84.83 & 32.35 & 37.09 \\
    - w attention & 80.77 & 79.24 & 23.77 & 26.54 \\
    \hdashline
    w/o soft sample &       &       &       &  \\
    - w uniform sample & 82.62 & 81.39 & 26.49 & 28.49 \\
    - w negative sample & 79.47 & 76.05 & 20.62 & 24.18 \\
    - w random sample & 82.03 & 80.88 & 27.50 & 29.73 \\
    \hdashline
    w/o aggregation &       &       &       &  \\
    - w/o $\mathcal{L}_n$  & 84.56 & 83.68 & 31.74 & 36.13 \\
    - w/o $\mathcal{L}_p$  & 84.88 & 83.97 & 30.87 & 34.48 \\
    - w/o $\mathcal{L}_p$ and $\mathcal{L}_n$ & 84.31 & 82.84 & 30.15 & 33.98 \\
    \midrule
    \midrule
    SAILER & -     & -     & 32.23 & 36.51 \\
    OPT-Match & 83.80 & 81.88 & 30.36 & 34.55 \\
    BERT  & 83.85 & 81.43 & 25.77 & 27.46 \\
    \bottomrule
    \end{tabular}%
    }
  \label{tab:addlabel}%
\end{table}%

\subsection{Ablation Study}
\label{sec:abla}

To verify the effect of the three steps of our learning framework, we perform an ablation study based on the BERT/SAILER + $SST^{ah}$. We remove the aggregation inference and find a mild decrease. This indicates that our framework benefits even for a single view. We remove the temporal aggregation strategy, which means we do not use any aggregation during training. We observe a significant decrease, indicating the importance of synthesizing representative views that contain various details. We also leverage the attention mechanism to aggregate these views as \cite{smith20cikm}. We observe an inferior performance even compared to the original BERT. This means that conventional methods are ineffective for some domain-specific tasks where redundant information must be further distinguished. 

We also test some variant view sample methods to validate the importance of choosing a salient aligned part. The random sample method means we directly sample continuous sentences to form a view, and it is not based on the subtopic analysis. The negative sampling only gets sentences from the small subtopic clusters, which is the opposite of the hard sampling. Their inferior performance indicates the importance of finding salient parts, as these sampling methods generally lack the primary aligned subtopics. We also give a further analysis of choosing a proper view in Section \ref{sec:furthertest}.

Finally, we test the adaptive cluster loss without the aggregation training. We find both losses contribute to the tasks. Most importantly, compared to other content selection methods, we notice that the vanilla-selected content may perform worse. This may be because the selected content is discontinuous, harming the knowledge learned in the pre-trained stage. 



\begin{figure}[t]
  \setlength{\belowcaptionskip}{-2pt} 
  \setlength{\abovecaptionskip}{-4pt}
  \includegraphics[width=\linewidth]{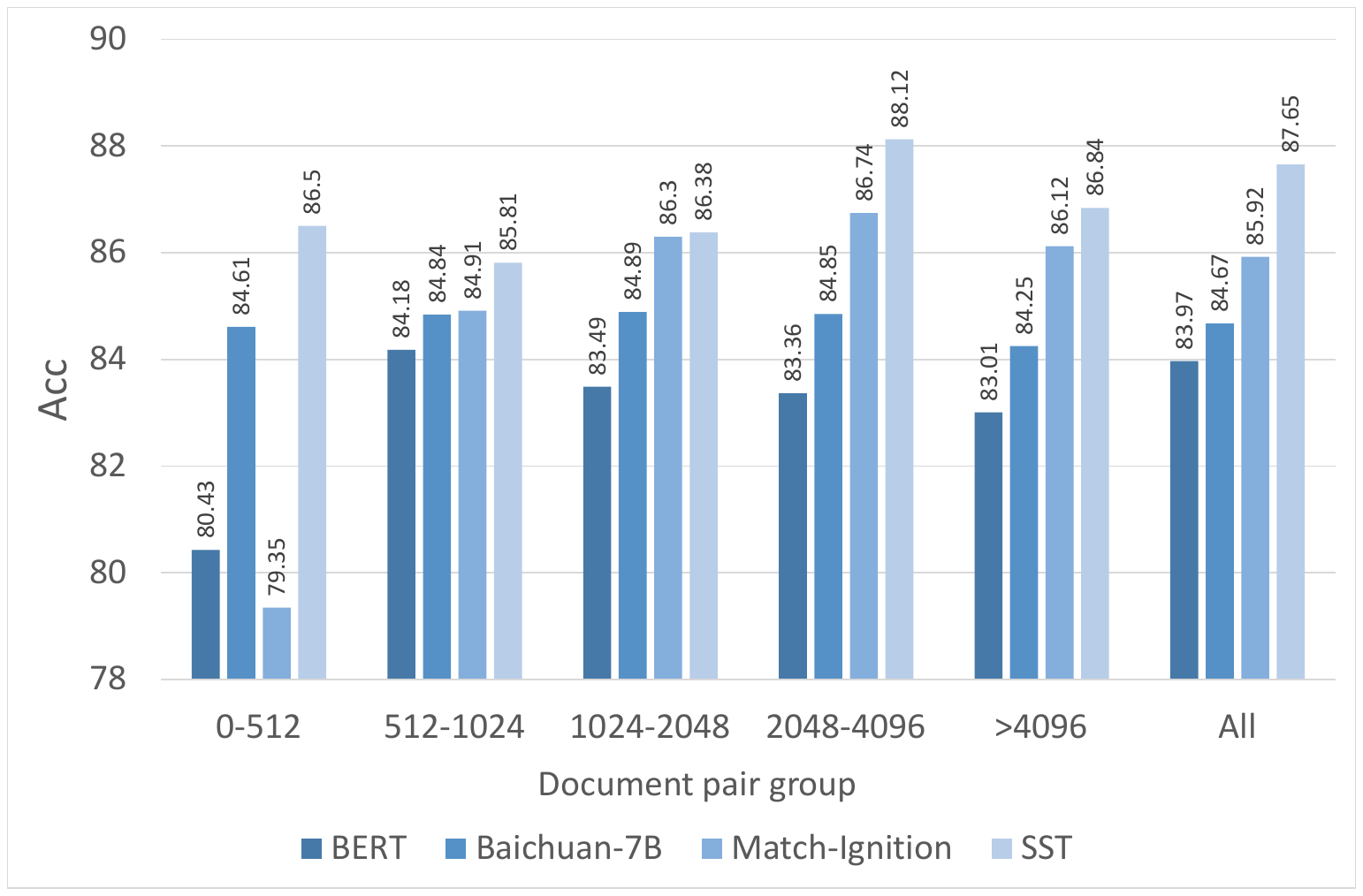}
  
  \caption{ Performance of typical models on document pairs of different lengths (\# words).}
  \label{fig: details result}
\end{figure}

\begin{figure}[t]
  \includegraphics[width=\linewidth]{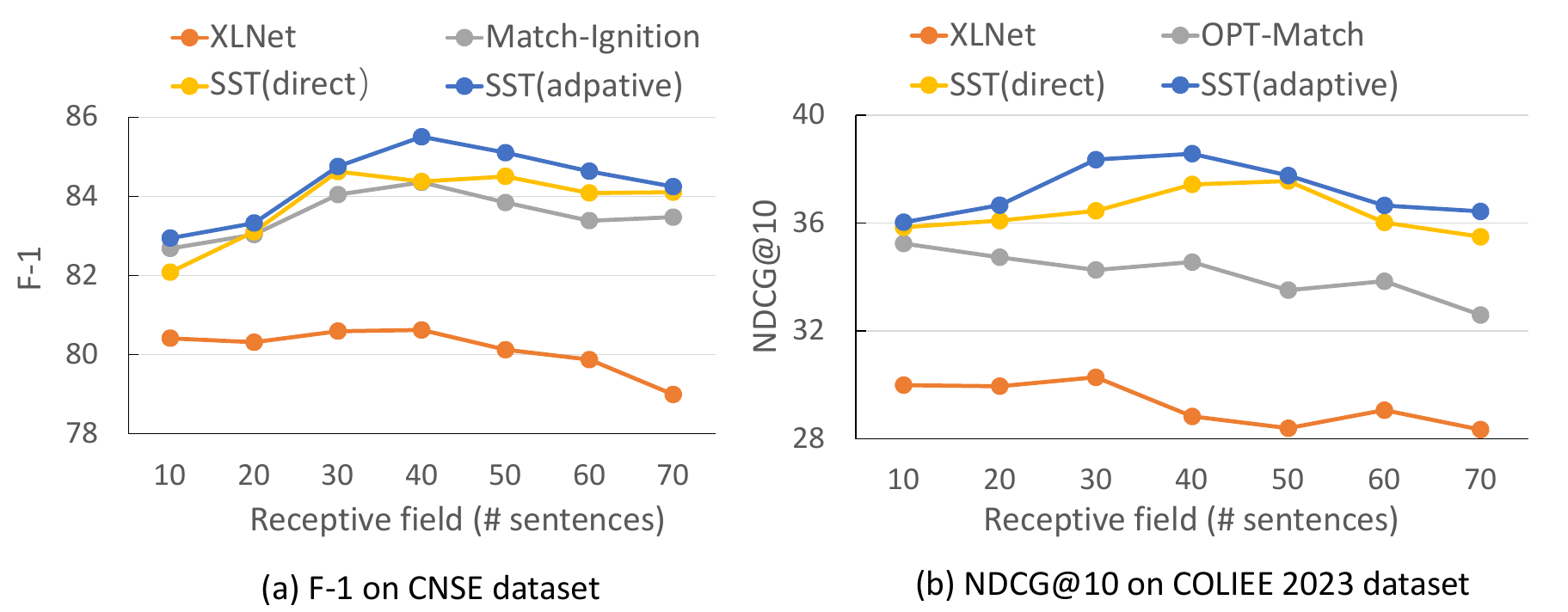}
  \caption{The impact of receptive fields of different model.}
  \label{fig: receptive_field}
\end{figure}

\subsection{Hyper-parameter Analysis}
\label{sec:hypertest}
\begin{figure}[t]
  \includegraphics[width=\linewidth]{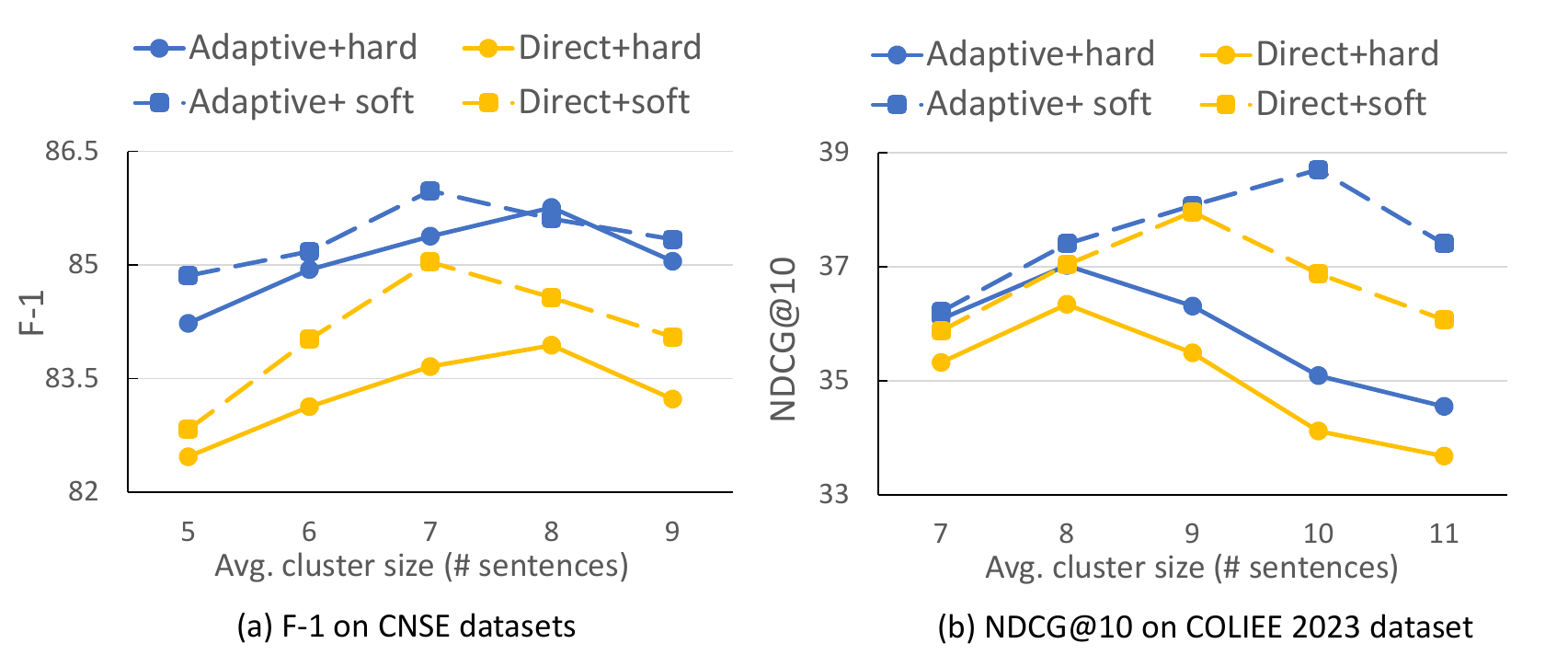}
  \caption{The impact of the average cluster size to SST framework.}
  \label{fig: cluster_size}
\end{figure}

To ensure comprehensive learning of the document, we need to control the receptive field of the SST framework. The results are reported in Figure \ref{fig: receptive_field}. Generally speaking, content selection models perform better than long-form models, and our framework performs better than content selection models. Moreover, adaptive clustering performs better than direct clustering, which shows better robustness as the confusing information increases.
Moreover, the quality of the subtopics is essential. The main hyper-parameter is the cluster numbers. The results are reported in Figure \ref{fig: cluster_size}. We vary the cluster size in a reasonable range and sampling with different views. We find that the soft sampling method is better and more stable than the hard sampling method. Some other hyper-parameters are discussed in Appendix \ref{App:hyper_test}.

\begin{figure}[t]
  \setlength{\belowcaptionskip}{-3pt} 
  \includegraphics[width=\linewidth]{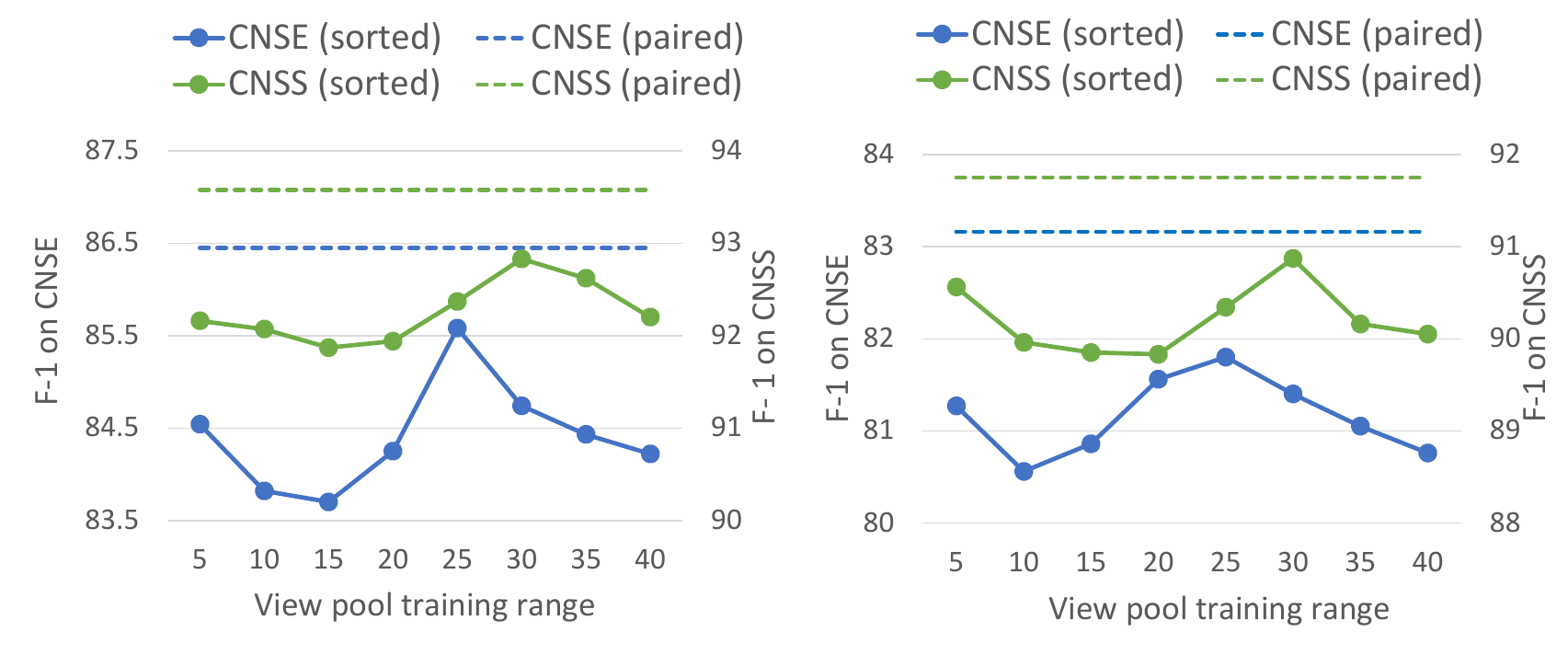}
  \caption{The influence of the heterogeneity and salience to the temporal aggregation.}
  \label{fig: temproal_aggregation}
\end{figure}

\subsection{Further Analysis}
\label{sec:furthertest}
According to previous discussion, representation has two dimensions: one is heterogeneity, and the other is aligned salience. We conduct further analysis to reveal how the heterogeneity and salience influence the temporal aggregation.
As we do not have the ground truth subtopic or alignment information, we leverage uniform, and soft sampling as proxy methods. We build a view pool and reorder the trained views. we sort the samples in descending order according to their similarity (use QL models \cite{statIR} here) to the original document for the random sample views. Generally, the more similar a view to the whole document, the more primary information it contains. Thus, training with these reordered views contains weaker aligned salience. Moreover, the top-ranked views have lower heterogeneity as they are similar to each other. We control the sampled range in the view pool. For example, if the view pool is 40 and the max training epoch is 4, we select the views from the range of $linspace(1,40,4)$, e.g., get 4 views from the range $[1,10],[11.20],[21,30],[31,40]$. In this experiment, we set the view pool size to 40 and the training epoch to 10. As the sampled range expands, the views have increasing heterogeneity and decreasing salience.

The results are shown in Figure \ref{fig: temproal_aggregation}. With these reordered samples, model perform worse than the original aligned ones. Interestingly, as the view pool increases, the curve of sampling shows the down-up-down trend. The first decrease may come from the homogeneity of views, similar to “spatial attention” The increase may come from the moderate heterogeneity and moderate alignment. The second decrease may come from the extra noise of these views as they neither align with each other nor contain primary information. Though the result come from a combination of factors, it indicates the importance of selecting representative views with both heterogeneity and salience. 

\begin{figure}[t]
  \includegraphics[width=\linewidth]{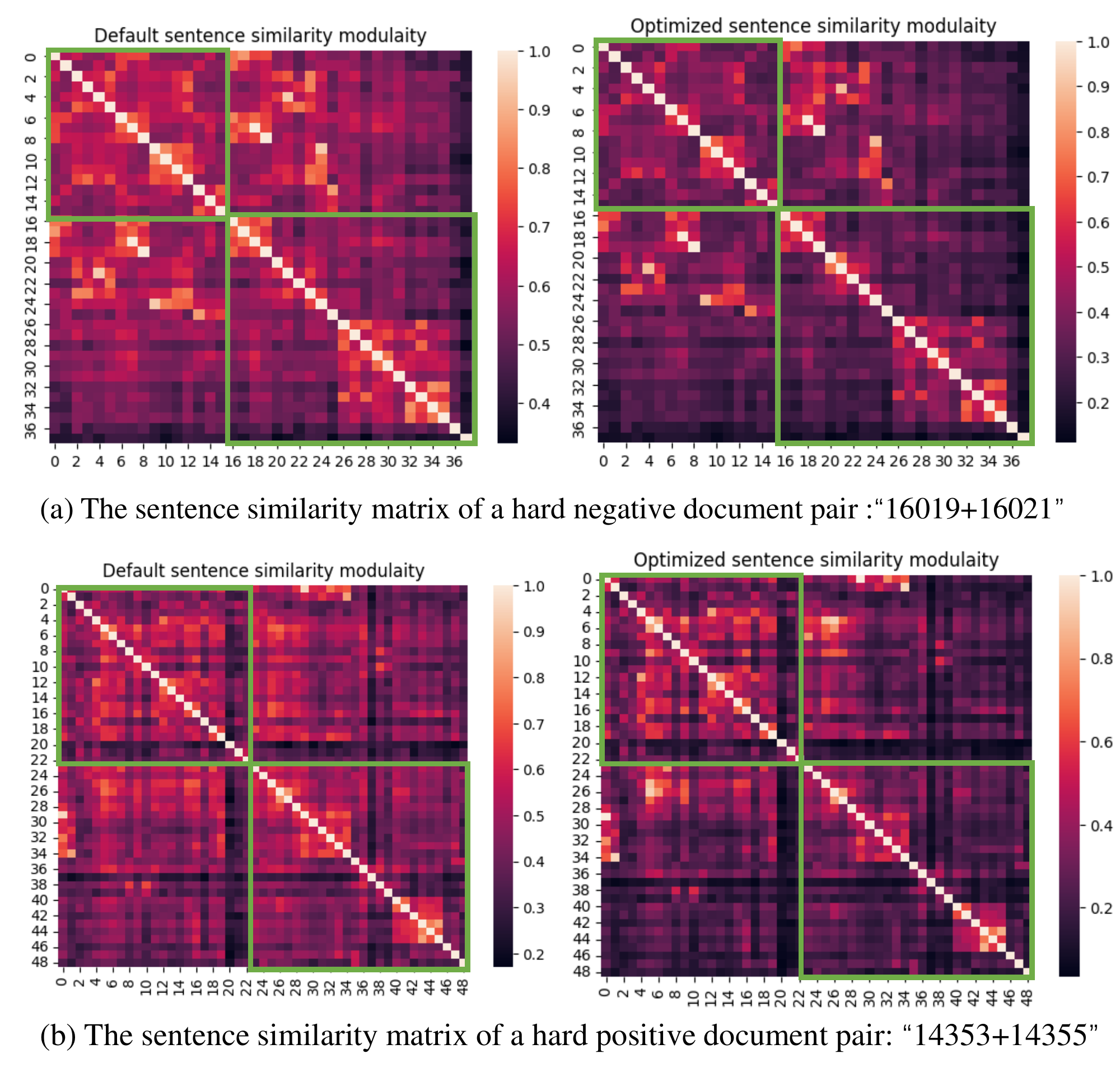}
  \caption{A case study of the optimized cluster. The similarity of sentences are measured with cosine similarity. The green square indicates the range of two documents correspondingly. }
  \label{fig: case}
\end{figure}

\subsection{Case Study}
\label{sec:case study}
To understand the modularity of document pairs, we illustrate the sentence relationship of documents before and with adaptive clustering. We give a sample of hard positive and negative samples of the CNSE dataset as in Figure \ref{fig: case}. The general similarity measure shows that the hard negative sample has many high-similarity sentences. After adaptive clustering optimization, the valuable aligned part becomes more distinct. The hard positive sample has vague similarity between documents; the optimized similarity shows better document alignment. Overall, the modularity information, including alignment and complementary, is further distinguished. 


%% file: sections/6_conclusion.tex
\section{Conclusion}

In this work, we propose a new learning framework SST for long-form document matching. We leverage the subtopic analysis to discover the modularity between documents, including the aligned and complementary information. These subtopics help us to find multiple matching signals between document. Then, we construct document views containing comprehensive and detailed valuable information. Furthermore, we propose a temporal aggregation strategy coupled with the view sampling strategy, which train the models efficiently and effectively. We conduct extensive experiments to investigate the effectiveness of different sampling methods and factors. These provide valuable insights into selecting an appropriate sampling strategy for the long-form document matching task. We also conduct further study to discuss the representativeness. We find multiple document parts containing heterogeneous and salient aligned details are representative for the task. Experimental results show that our training frameworks are effective on multiple long-form document matching tasks. 

\clearpage

%% file: sections/appendix.tex
\section{Baseline models}
\label{App:baseline models}

\begin{itemize}
\item Lexical based models. Traditional term-based methods still act as a baseline in retrieval, especially in legal case retrieval, which has no input limits. BM25 \cite{bm25}. Meanwhile, the BERTTopc \cite{berttopic} model is an unsupervised model used to get the latent subtopics and classify them.

\item Hierarchical models: Most hierarchical models are passage-level interaction models. Passage-level interaction models mean splitting original documents into multiple passages, leveraging pre-trained models to get representations, and then aggregating passage representations to get the final results. As breaking a whole document into passages leads to less interaction among passages, these models generally serve as efficient recall models  \cite{cda20emnlp,smith20cikm} or require additional supervision like Bert-pli  \cite{bertpli20ijcai}. Here, we adopt BERT-PLI in supervised training settings as a reference since it achieved second place in the COLIEE2020 competition.

\item Long context models: Long context models are those who have longer input as BERT. For example, the input size of XLNET \cite{xlnet19nips} and lawformer \cite{lawfm21} are both 4096 tokens. Recently, LLMs have also been used for ranking tasks  \cite{rankllama23sigir}. We leverage Chinese LLMs Baichuan-7B \cite{baichuan23}, whose input limit is 2048 tokens. We conduct RankT5 \cite{rankt5_23sigir} style training for news duplication tasks. 

\item Content selection models:  Content selection models, including truncated models and selection models. A truncated model means directly truncating a static part for models to train and infer. For the pre-trained model, we include BERT cross-encoder and Match-Ignition(denoted as Match-Ig). Match-Ignition is the previous SOTA model for news duplication tasks. We also adopt the structure-aware pre-trained model SAILER  \cite{sailer23sigir} on COLIEE as it is a first-stage model and is not suitable for reranking. Graph models are also content selection methods that select sentences as node representations. The CIG model \cite{cig19acl} is specially designed for news duplication tasks, which proposes to leverage the critical named entity to represent that concept with both lexical and dense representation to aggregate with GCN models. As the model needs to identify the salient concept, such as a named entity, we do not apply it to legal retrieval.
\end{itemize}

\section{Implementation Details}
\label{App:implementation details}

\subsection{Training Configuration}
We implement the clustering step of the subtopic-aware view sampling strategy with off-the-shelf tools or models. In the direct clustering step, we use the publicly available tool nltk \footnote{https://www.nltk.org/} to split the document into sentences and do lemmatization for English documents when building a similarity matrix. We use the sklearn tools to perform clustering and set the cluster number according to the expected cluster size of 6. 
For the adaptive clustering, the original implementation is based on the GCN network, which is suited for a big and static graph, which is not suitable for our scenery. We modified it to a GAT network, which is suitable for sentence similarity graphs of different sizes. We also process the adjacency matrix with normalization and sparsification as \cite{sdcn20www}. We control the positive and negative samples to be balanced and set the static learning rate as 1e-3. 
For matching models, we set the receptive field of models to 40 sentences. We leverage the Adam optimizer and linear-up scheduler to train our model and set the learning rate to 1e-5. For the news duplication tasks, we use the cross-encoder structure training with cross entropy. For Bert-based models, we limit the input of the query and the candidate documents to the length of 256 tokens. We keep our model with the same view size according to the average sentence length. For legal case retrieval, we use the dual-encoder structure trained with InfoNCE loss. For Bert-based models, we limit the input of the query and the candidate documents to the length of 512 tokens. In the aggregation inference stage, we set the view pool size as 3.

\subsection{Prompts of GPT models}

As it is hard to choose a proper prompt and demonstration for in-context learning, the experiments of GPT models are conducted in the zero-shot learning way. The prompts (translated from Chinese) are organized as follows for news duplication tasks:

\mdfsetup{backgroundcolor=gray!10}

\begin{mdframed}	
Please analysis whether the two documents listed below tell the same news events. Notice that only answer “yes” or ”no“ and do not give other utterence. Document 1: [content of Document 1]  Document 2  [content of Document 2].
\end{mdframed}

\section{Hyper-parameters test}
\label{App:hyper_test}

We conduct a hyper-parameter test of $\lambda$ of the unsupervised model and test the effect for soft sampling. Both the $\mathcal{L}_p$ and $\mathcal{L}_n$ contribute to the unsupervised model. We find that the general performance of the unsupervised model is robust to changes in $\lambda$ values as it changes from 0.02 to 50.

\begin{figure}[t]
  \includegraphics[width=\linewidth]{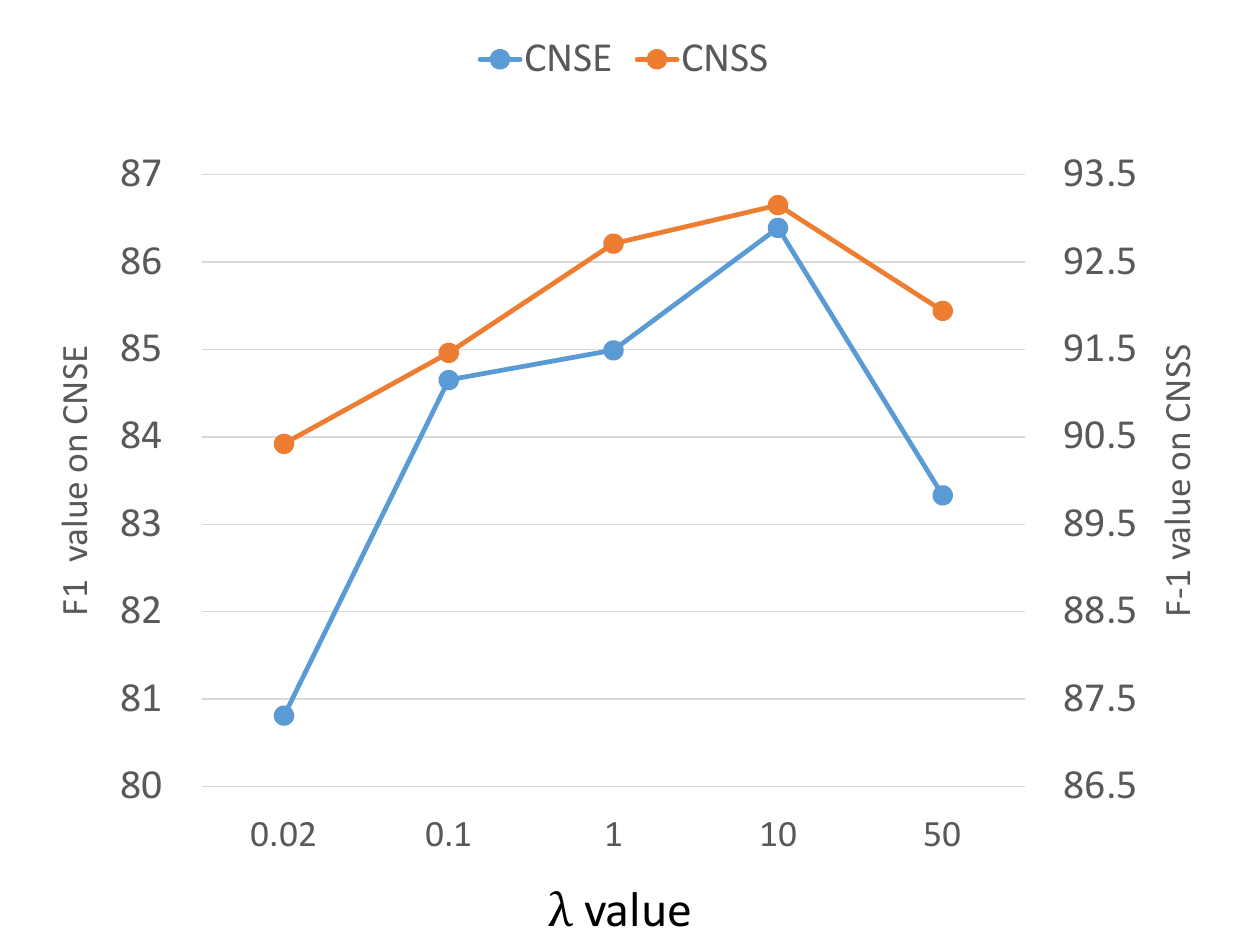}
  \caption{The effect of $\lambda$ for training unsupervised clustering metrics.}
  \label{fig: lambda}
\end{figure}

%% file: sigconf-sample.bbl

\begin{thebibliography}{52}


\ifx \showCODEN    \undefined \def \showCODEN     #1{\unskip}     \fi
\ifx \showDOI      \undefined \def \showDOI       #1{#1}\fi
\ifx \showISBNx    \undefined \def \showISBNx     #1{\unskip}     \fi
\ifx \showISBNxiii \undefined \def \showISBNxiii  #1{\unskip}     \fi
\ifx \showISSN     \undefined \def \showISSN      #1{\unskip}     \fi
\ifx \showLCCN     \undefined \def \showLCCN      #1{\unskip}     \fi
\ifx \shownote     \undefined \def \shownote      #1{#1}          \fi
\ifx \showarticletitle \undefined \def \showarticletitle #1{#1}   \fi
\ifx \showURL      \undefined \def \showURL       {\relax}        \fi
\providecommand\bibfield[2]{#2}
\providecommand\bibinfo[2]{#2}
\providecommand\natexlab[1]{#1}
\providecommand\showeprint[2][]{arXiv:#2}

\bibitem[Althammer et~al\mbox{.}(2022)]%
        {parm22ecir}
\bibfield{author}{\bibinfo{person}{Sophia Althammer}, \bibinfo{person}{Sebastian Hofst{\"a}tter}, \bibinfo{person}{Mete Sertkan}, \bibinfo{person}{Suzan Verberne}, {and} \bibinfo{person}{Allan Hanbury}.} \bibinfo{year}{2022}\natexlab{}.
\newblock \showarticletitle{PARM: A Paragraph Aggregation Retrieval Model for Dense Document-to-Document Retrieval}. In \bibinfo{booktitle}{\emph{Advances in Information Retrieval}}, \bibfield{editor}{\bibinfo{person}{Matthias Hagen}, \bibinfo{person}{Suzan Verberne}, \bibinfo{person}{Craig Macdonald}, \bibinfo{person}{Christin Seifert}, \bibinfo{person}{Krisztian Balog}, \bibinfo{person}{Kjetil N{\o}rv{\aa}g}, {and} \bibinfo{person}{Vinay Setty}} (Eds.). \bibinfo{publisher}{Springer International Publishing}, \bibinfo{address}{Cham}, \bibinfo{pages}{19--34}.
\newblock
\showISBNx{978-3-030-99736-6}


\bibitem[Arnold et~al\mbox{.}(2019)]%
        {sector19tacl}
\bibfield{author}{\bibinfo{person}{Sebastian Arnold}, \bibinfo{person}{Rudolf Schneider}, \bibinfo{person}{Philippe Cudr{\'e}-Mauroux}, \bibinfo{person}{Felix~A Gers}, {and} \bibinfo{person}{Alexander L{\"o}ser}.} \bibinfo{year}{2019}\natexlab{}.
\newblock \showarticletitle{SECTOR: A neural model for coherent topic segmentation and classification}.
\newblock \bibinfo{journal}{\emph{Transactions of the Association for Computational Linguistics}}  \bibinfo{volume}{7} (\bibinfo{year}{2019}), \bibinfo{pages}{169--184}.
\newblock


\bibitem[Askari et~al\mbox{.}(2024)]%
        {rprs24tois}
\bibfield{author}{\bibinfo{person}{Arian Askari}, \bibinfo{person}{Suzan Verberne}, \bibinfo{person}{Amin Abolghasemi}, \bibinfo{person}{Wessel Kraaij}, {and} \bibinfo{person}{Gabriella Pasi}.} \bibinfo{year}{2024}\natexlab{}.
\newblock \showarticletitle{Retrieval for Extremely Long Queries and Documents with {RPRS:} {A} Highly Efficient and Effective Transformer-based Re-Ranker}.
\newblock \bibinfo{journal}{\emph{{ACM} Trans. Inf. Syst.}} \bibinfo{volume}{42}, \bibinfo{number}{5} (\bibinfo{year}{2024}), \bibinfo{pages}{115:1--115:32}.
\newblock
\urldef\tempurl%
\url{https://doi.org/10.1145/3631938}
\showDOI{\tempurl}


\bibitem[Bai et~al\mbox{.}(2023)]%
        {segformer23aaai}
\bibfield{author}{\bibinfo{person}{Haitao Bai}, \bibinfo{person}{Pinghui Wang}, \bibinfo{person}{Ruofei Zhang}, {and} \bibinfo{person}{Zhou Su}.} \bibinfo{year}{2023}\natexlab{}.
\newblock \showarticletitle{SegFormer: a topic segmentation model with controllable range of attention}. In \bibinfo{booktitle}{\emph{Proceedings of the Thirty-Seventh AAAI Conference on Artificial Intelligence and Thirty-Fifth Conference on Innovative Applications of Artificial Intelligence and Thirteenth Symposium on Educational Advances in Artificial Intelligence}} \emph{(\bibinfo{series}{AAAI'23/IAAI'23/EAAI'23})}. \bibinfo{publisher}{AAAI Press}, Article \bibinfo{articleno}{1408}, \bibinfo{numpages}{8}~pages.
\newblock
\showISBNx{978-1-57735-880-0}
\urldef\tempurl%
\url{https://doi.org/10.1609/aaai.v37i11.26477}
\showDOI{\tempurl}


\bibitem[Bianchi et~al\mbox{.}(2020)]%
        {unspecral20icml}
\bibfield{author}{\bibinfo{person}{Filippo~Maria Bianchi}, \bibinfo{person}{Daniele Grattarola}, {and} \bibinfo{person}{Cesare Alippi}.} \bibinfo{year}{2020}\natexlab{}.
\newblock \showarticletitle{Spectral clustering with graph neural networks for graph pooling}. In \bibinfo{booktitle}{\emph{International conference on machine learning}}. PMLR, \bibinfo{pages}{874--883}.
\newblock


\bibitem[Bo et~al\mbox{.}(2020)]%
        {sdcn20www}
\bibfield{author}{\bibinfo{person}{Deyu Bo}, \bibinfo{person}{Xiao Wang}, \bibinfo{person}{Chuan Shi}, \bibinfo{person}{Meiqi Zhu}, \bibinfo{person}{Emiao Lu}, {and} \bibinfo{person}{Peng Cui}.} \bibinfo{year}{2020}\natexlab{}.
\newblock \showarticletitle{Structural deep clustering network}. In \bibinfo{booktitle}{\emph{Proceedings of the web conference 2020}}. \bibinfo{pages}{1400--1410}.
\newblock


\bibitem[Deng et~al\mbox{.}(2024)]%
        {multi_grain_srd24tois}
\bibfield{author}{\bibinfo{person}{Zhirui Deng}, \bibinfo{person}{Zhicheng Dou}, \bibinfo{person}{Zhan Su}, {and} \bibinfo{person}{Ji-Rong Wen}.} \bibinfo{year}{2024}\natexlab{}.
\newblock \showarticletitle{Multi-grained Document Modeling for Search Result Diversification}.
\newblock \bibinfo{journal}{\emph{ACM Transactions on Information Systems}} \bibinfo{volume}{42}, \bibinfo{number}{5} (\bibinfo{year}{2024}), \bibinfo{pages}{1--22}.
\newblock


\bibitem[Devlin et~al\mbox{.}(2019)]%
        {bert}
\bibfield{author}{\bibinfo{person}{Jacob Devlin}, \bibinfo{person}{Ming-Wei Chang}, \bibinfo{person}{Kenton Lee}, {and} \bibinfo{person}{Kristina Toutanova}.} \bibinfo{year}{2019}\natexlab{}.
\newblock \showarticletitle{BERT: Pre-training of Deep Bidirectional Transformers for Language Understanding}. In \bibinfo{booktitle}{\emph{Proceedings of the 2019 Conference of the North American Chapter of the Association for Computational Linguistics: Human Language Technologies, Volume 1 (Long and Short Papers)}}. \bibinfo{pages}{4171--4186}.
\newblock


\bibitem[Dong et~al\mbox{.}(2024)]%
        {rlcf24sigir}
\bibfield{author}{\bibinfo{person}{Qian Dong}, \bibinfo{person}{Yiding Liu}, \bibinfo{person}{Qingyao Ai}, \bibinfo{person}{Zhijing Wu}, \bibinfo{person}{Haitao Li}, \bibinfo{person}{Yiqun Liu}, \bibinfo{person}{Shuaiqiang Wang}, \bibinfo{person}{Dawei Yin}, {and} \bibinfo{person}{Shaoping Ma}.} \bibinfo{year}{2024}\natexlab{}.
\newblock \showarticletitle{Unsupervised Large Language Model Alignment for Information Retrieval via Contrastive Feedback}. In \bibinfo{booktitle}{\emph{Proceedings of the 47th International ACM SIGIR Conference on Research and Development in Information Retrieval}} (Washington DC, USA) \emph{(\bibinfo{series}{SIGIR '24})}. \bibinfo{publisher}{Association for Computing Machinery}, \bibinfo{address}{New York, NY, USA}, \bibinfo{pages}{48–58}.
\newblock
\showISBNx{9798400704314}
\urldef\tempurl%
\url{https://doi.org/10.1145/3626772.3657689}
\showDOI{\tempurl}


\bibitem[Ginzburg et~al\mbox{.}(2021)]%
        {sdr21acl_finding}
\bibfield{author}{\bibinfo{person}{Dvir Ginzburg}, \bibinfo{person}{Itzik Malkiel}, \bibinfo{person}{Oren Barkan}, \bibinfo{person}{Avi Caciularu}, {and} \bibinfo{person}{Noam Koenigstein}.} \bibinfo{year}{2021}\natexlab{}.
\newblock \showarticletitle{Self-Supervised Document Similarity Ranking via Contextualized Language Models and Hierarchical Inference}. In \bibinfo{booktitle}{\emph{Findings of the Association for Computational Linguistics: {ACL/IJCNLP} 2021, Online Event, August 1-6, 2021}} \emph{(\bibinfo{series}{Findings of {ACL}}, Vol.~\bibinfo{volume}{{ACL/IJCNLP} 2021})}, \bibfield{editor}{\bibinfo{person}{Chengqing Zong}, \bibinfo{person}{Fei Xia}, \bibinfo{person}{Wenjie Li}, {and} \bibinfo{person}{Roberto Navigli}} (Eds.). \bibinfo{publisher}{Association for Computational Linguistics}, \bibinfo{pages}{3088--3098}.
\newblock
\urldef\tempurl%
\url{https://doi.org/10.18653/V1/2021.FINDINGS-ACL.272}
\showDOI{\tempurl}


\bibitem[Glava{\v{s}} et~al\mbox{.}(2016)]%
        {graphseg16acl}
\bibfield{author}{\bibinfo{person}{Goran Glava{\v{s}}}, \bibinfo{person}{Federico Nanni}, {and} \bibinfo{person}{Simone~Paolo Ponzetto}.} \bibinfo{year}{2016}\natexlab{}.
\newblock \showarticletitle{Unsupervised text segmentation using semantic relatedness graphs}. In \bibinfo{booktitle}{\emph{Proceedings of the Fifth Joint Conference on Lexical and Computational Semantics}}. Association for Computational Linguistics, \bibinfo{pages}{125--130}.
\newblock


\bibitem[Goebel et~al\mbox{.}(2023)]%
        {coliee23summary}
\bibfield{author}{\bibinfo{person}{Randy Goebel}, \bibinfo{person}{Yoshinobu Kano}, \bibinfo{person}{Mi-Young Kim}, \bibinfo{person}{Juliano Rabelo}, \bibinfo{person}{Ken Satoh}, {and} \bibinfo{person}{Masaharu Yoshioka}.} \bibinfo{year}{2023}\natexlab{}.
\newblock \showarticletitle{Summary of the Competition on Legal Information, Extraction/Entailment (COLIEE) 2023}. In \bibinfo{booktitle}{\emph{Proceedings of the Nineteenth International Conference on Artificial Intelligence and Law}}. \bibinfo{pages}{472--480}.
\newblock


\bibitem[Grootendorst(2022)]%
        {berttopic}
\bibfield{author}{\bibinfo{person}{Maarten Grootendorst}.} \bibinfo{year}{2022}\natexlab{}.
\newblock \showarticletitle{BERTopic: Neural topic modeling with a class-based TF-IDF procedure}.
\newblock \bibinfo{journal}{\emph{arXiv preprint arXiv:2203.05794}} (\bibinfo{year}{2022}).
\newblock


\bibitem[Hearst and Plaunt(1993)]%
        {texttilling}
\bibfield{author}{\bibinfo{person}{Marti~A Hearst} {and} \bibinfo{person}{Christian Plaunt}.} \bibinfo{year}{1993}\natexlab{}.
\newblock \showarticletitle{Subtopic structuring for full-length document access}. In \bibinfo{booktitle}{\emph{Proceedings of the 16th annual international ACM SIGIR conference on Research and development in information retrieval}}. \bibinfo{pages}{59--68}.
\newblock


\bibitem[Hofst{\"{a}}tter et~al\mbox{.}(2021)]%
        {Intra-select21sigir}
\bibfield{author}{\bibinfo{person}{Sebastian Hofst{\"{a}}tter}, \bibinfo{person}{Bhaskar Mitra}, \bibinfo{person}{Hamed Zamani}, \bibinfo{person}{Nick Craswell}, {and} \bibinfo{person}{Allan Hanbury}.} \bibinfo{year}{2021}\natexlab{}.
\newblock \showarticletitle{Intra-Document Cascading: Learning to Select Passages for Neural Document Ranking}. In \bibinfo{booktitle}{\emph{{SIGIR} '21: The 44th International {ACM} {SIGIR} Conference on Research and Development in Information Retrieval, Virtual Event, Canada, July 11-15, 2021}}, \bibfield{editor}{\bibinfo{person}{Fernando Diaz}, \bibinfo{person}{Chirag Shah}, \bibinfo{person}{Torsten Suel}, \bibinfo{person}{Pablo Castells}, \bibinfo{person}{Rosie Jones}, {and} \bibinfo{person}{Tetsuya Sakai}} (Eds.). \bibinfo{publisher}{{ACM}}, \bibinfo{pages}{1349--1358}.
\newblock
\urldef\tempurl%
\url{https://doi.org/10.1145/3404835.3462889}
\showDOI{\tempurl}


\bibitem[Jha et~al\mbox{.}(2023)]%
        {colde23tkdd}
\bibfield{author}{\bibinfo{person}{Akshita Jha}, \bibinfo{person}{Vineeth Rakesh}, \bibinfo{person}{Jaideep Chandrashekar}, \bibinfo{person}{Adithya Samavedhi}, {and} \bibinfo{person}{Chandan~K. Reddy}.} \bibinfo{year}{2023}\natexlab{}.
\newblock \showarticletitle{Supervised Contrastive Learning for Interpretable Long-Form Document Matching}.
\newblock \bibinfo{journal}{\emph{{ACM} Trans. Knowl. Discov. Data}} \bibinfo{volume}{17}, \bibinfo{number}{2} (\bibinfo{year}{2023}), \bibinfo{pages}{27:1--27:17}.
\newblock
\urldef\tempurl%
\url{https://doi.org/10.1145/3542822}
\showDOI{\tempurl}


\bibitem[Jiang et~al\mbox{.}(2019)]%
        {smash19www}
\bibfield{author}{\bibinfo{person}{Jyun{-}Yu Jiang}, \bibinfo{person}{Mingyang Zhang}, \bibinfo{person}{Cheng Li}, \bibinfo{person}{Michael Bendersky}, \bibinfo{person}{Nadav Golbandi}, {and} \bibinfo{person}{Marc Najork}.} \bibinfo{year}{2019}\natexlab{}.
\newblock \showarticletitle{Semantic Text Matching for Long-Form Documents}. In \bibinfo{booktitle}{\emph{The World Wide Web Conference, {WWW} 2019, San Francisco, CA, USA, May 13-17, 2019}}, \bibfield{editor}{\bibinfo{person}{Ling Liu}, \bibinfo{person}{Ryen~W. White}, \bibinfo{person}{Amin Mantrach}, \bibinfo{person}{Fabrizio Silvestri}, \bibinfo{person}{Julian~J. McAuley}, \bibinfo{person}{Ricardo Baeza{-}Yates}, {and} \bibinfo{person}{Leila Zia}} (Eds.). \bibinfo{publisher}{{ACM}}, \bibinfo{pages}{795--806}.
\newblock
\urldef\tempurl%
\url{https://doi.org/10.1145/3308558.3313707}
\showDOI{\tempurl}


\bibitem[Leonhardt et~al\mbox{.}(2023)]%
        {exdocrank23tois}
\bibfield{author}{\bibinfo{person}{Jurek Leonhardt}, \bibinfo{person}{Koustav Rudra}, {and} \bibinfo{person}{Avishek Anand}.} \bibinfo{year}{2023}\natexlab{}.
\newblock \showarticletitle{Extractive explanations for interpretable text ranking}.
\newblock \bibinfo{journal}{\emph{ACM Transactions on Information Systems}} \bibinfo{volume}{41}, \bibinfo{number}{4} (\bibinfo{year}{2023}), \bibinfo{pages}{1--31}.
\newblock


\bibitem[Li et~al\mbox{.}(2023)]%
        {sailer23sigir}
\bibfield{author}{\bibinfo{person}{Haitao Li}, \bibinfo{person}{Qingyao Ai}, \bibinfo{person}{Jia Chen}, \bibinfo{person}{Qian Dong}, \bibinfo{person}{Yueyue Wu}, \bibinfo{person}{Yiqun Liu}, \bibinfo{person}{Chong Chen}, {and} \bibinfo{person}{Qi Tian}.} \bibinfo{year}{2023}\natexlab{}.
\newblock \showarticletitle{{SAILER:} Structure-aware Pre-trained Language Model for Legal Case Retrieval}. In \bibinfo{booktitle}{\emph{Proceedings of the 46th International {ACM} {SIGIR} Conference on Research and Development in Information Retrieval, {SIGIR} 2023, Taipei, Taiwan, July 23-27, 2023}}, \bibfield{editor}{\bibinfo{person}{Hsin{-}Hsi Chen}, \bibinfo{person}{Wei{-}Jou~(Edward) Duh}, \bibinfo{person}{Hen{-}Hsen Huang}, \bibinfo{person}{Makoto~P. Kato}, \bibinfo{person}{Josiane Mothe}, {and} \bibinfo{person}{Barbara Poblete}} (Eds.). \bibinfo{publisher}{{ACM}}, \bibinfo{pages}{1035--1044}.
\newblock
\urldef\tempurl%
\url{https://doi.org/10.1145/3539618.3591761}
\showDOI{\tempurl}


\bibitem[Li et~al\mbox{.}(2024)]%
        {lecardv2_24sigir}
\bibfield{author}{\bibinfo{person}{Haitao Li}, \bibinfo{person}{Yunqiu Shao}, \bibinfo{person}{Yueyue Wu}, \bibinfo{person}{Qingyao Ai}, \bibinfo{person}{Yixiao Ma}, {and} \bibinfo{person}{Yiqun Liu}.} \bibinfo{year}{2024}\natexlab{}.
\newblock \showarticletitle{LeCaRDv2: A Large-Scale Chinese Legal Case Retrieval Dataset}. In \bibinfo{booktitle}{\emph{Proceedings of the 47th International ACM SIGIR Conference on Research and Development in Information Retrieval}}. \bibinfo{pages}{2251--2260}.
\newblock


\bibitem[Lipowski and Lipowska(2012)]%
        {Roulette-wheel}
\bibfield{author}{\bibinfo{person}{Adam Lipowski} {and} \bibinfo{person}{Dorota Lipowska}.} \bibinfo{year}{2012}\natexlab{}.
\newblock \showarticletitle{Roulette-wheel selection via stochastic acceptance}.
\newblock \bibinfo{journal}{\emph{Physica A: Statistical Mechanics and its Applications}} \bibinfo{volume}{391}, \bibinfo{number}{6} (\bibinfo{year}{2012}), \bibinfo{pages}{2193--2196}.
\newblock


\bibitem[Liu et~al\mbox{.}(2019)]%
        {cig19acl}
\bibfield{author}{\bibinfo{person}{Bang Liu}, \bibinfo{person}{Di Niu}, \bibinfo{person}{Haojie Wei}, \bibinfo{person}{Jinghong Lin}, \bibinfo{person}{Yancheng He}, \bibinfo{person}{Kunfeng Lai}, {and} \bibinfo{person}{Yu Xu}.} \bibinfo{year}{2019}\natexlab{}.
\newblock \showarticletitle{Matching Article Pairs with Graphical Decomposition and Convolutions}. In \bibinfo{booktitle}{\emph{Proceedings of the 57th Conference of the Association for Computational Linguistics, {ACL} 2019, Florence, Italy, July 28- August 2, 2019, Volume 1: Long Papers}}, \bibfield{editor}{\bibinfo{person}{Anna Korhonen}, \bibinfo{person}{David~R. Traum}, {and} \bibinfo{person}{Llu{\'{\i}}s M{\`{a}}rquez}} (Eds.). \bibinfo{publisher}{Association for Computational Linguistics}, \bibinfo{pages}{6284--6294}.
\newblock
\urldef\tempurl%
\url{https://doi.org/10.18653/V1/P19-1632}
\showDOI{\tempurl}


\bibitem[Ma et~al\mbox{.}(2024)]%
        {rankllama23sigir}
\bibfield{author}{\bibinfo{person}{Xueguang Ma}, \bibinfo{person}{Liang Wang}, \bibinfo{person}{Nan Yang}, \bibinfo{person}{Furu Wei}, {and} \bibinfo{person}{Jimmy Lin}.} \bibinfo{year}{2024}\natexlab{}.
\newblock \showarticletitle{Fine-tuning llama for multi-stage text retrieval}. In \bibinfo{booktitle}{\emph{Proceedings of the 47th International ACM SIGIR Conference on Research and Development in Information Retrieval}}. \bibinfo{pages}{2421--2425}.
\newblock


\bibitem[Ma et~al\mbox{.}(2023)]%
        {slr23tois}
\bibfield{author}{\bibinfo{person}{Yixiao Ma}, \bibinfo{person}{Yueyue Wu}, \bibinfo{person}{Qingyao Ai}, \bibinfo{person}{Yiqun Liu}, \bibinfo{person}{Yunqiu Shao}, \bibinfo{person}{Min Zhang}, {and} \bibinfo{person}{Shaoping Ma}.} \bibinfo{year}{2023}\natexlab{}.
\newblock \showarticletitle{Incorporating Structural Information into Legal Case Retrieval}.
\newblock \bibinfo{journal}{\emph{ACM Transactions on Information Systems}} \bibinfo{volume}{42}, \bibinfo{number}{2} (\bibinfo{year}{2023}), \bibinfo{pages}{1--28}.
\newblock


\bibitem[Mihalcea and Tarau(2004)]%
        {textrank}
\bibfield{author}{\bibinfo{person}{Rada Mihalcea} {and} \bibinfo{person}{Paul Tarau}.} \bibinfo{year}{2004}\natexlab{}.
\newblock \showarticletitle{Textrank: Bringing order into text}. In \bibinfo{booktitle}{\emph{Proceedings of the 2004 conference on empirical methods in natural language processing}}. \bibinfo{pages}{404--411}.
\newblock


\bibitem[Ostendorff et~al\mbox{.}(2022)]%
        {ncl22emnlp}
\bibfield{author}{\bibinfo{person}{Malte Ostendorff}, \bibinfo{person}{Nils Rethmeier}, \bibinfo{person}{Isabelle Augenstein}, \bibinfo{person}{Bela Gipp}, {and} \bibinfo{person}{Georg Rehm}.} \bibinfo{year}{2022}\natexlab{}.
\newblock \showarticletitle{Neighborhood Contrastive Learning for Scientific Document Representations with Citation Embeddings}. In \bibinfo{booktitle}{\emph{Proceedings of the 2022 Conference on Empirical Methods in Natural Language Processing, {EMNLP} 2022, Abu Dhabi, United Arab Emirates, December 7-11, 2022}}, \bibfield{editor}{\bibinfo{person}{Yoav Goldberg}, \bibinfo{person}{Zornitsa Kozareva}, {and} \bibinfo{person}{Yue Zhang}} (Eds.). \bibinfo{publisher}{Association for Computational Linguistics}, \bibinfo{pages}{11670--11688}.
\newblock
\urldef\tempurl%
\url{https://doi.org/10.18653/V1/2022.EMNLP-MAIN.802}
\showDOI{\tempurl}


\bibitem[Pang et~al\mbox{.}(2021)]%
        {match_ig21cikm}
\bibfield{author}{\bibinfo{person}{Liang Pang}, \bibinfo{person}{Yanyan Lan}, {and} \bibinfo{person}{Xueqi Cheng}.} \bibinfo{year}{2021}\natexlab{}.
\newblock \showarticletitle{Match-Ignition: Plugging PageRank into Transformer for Long-form Text Matching}. In \bibinfo{booktitle}{\emph{{CIKM} '21: The 30th {ACM} International Conference on Information and Knowledge Management, Virtual Event, Queensland, Australia, November 1 - 5, 2021}}, \bibfield{editor}{\bibinfo{person}{Gianluca Demartini}, \bibinfo{person}{Guido Zuccon}, \bibinfo{person}{J.~Shane Culpepper}, \bibinfo{person}{Zi~Huang}, {and} \bibinfo{person}{Hanghang Tong}} (Eds.). \bibinfo{publisher}{{ACM}}, \bibinfo{pages}{1396--1405}.
\newblock
\urldef\tempurl%
\url{https://doi.org/10.1145/3459637.3482450}
\showDOI{\tempurl}


\bibitem[Riedl and Biemann(2012)]%
        {topictiling}
\bibfield{author}{\bibinfo{person}{Martin Riedl} {and} \bibinfo{person}{Chris Biemann}.} \bibinfo{year}{2012}\natexlab{}.
\newblock \showarticletitle{TopicTiling: a text segmentation algorithm based on LDA}. In \bibinfo{booktitle}{\emph{Proceedings of ACL 2012 student research workshop}}. \bibinfo{pages}{37--42}.
\newblock


\bibitem[Robertson et~al\mbox{.}(2009)]%
        {bm25}
\bibfield{author}{\bibinfo{person}{Stephen Robertson}, \bibinfo{person}{Hugo Zaragoza}, {et~al\mbox{.}}} \bibinfo{year}{2009}\natexlab{}.
\newblock \showarticletitle{The probabilistic relevance framework: BM25 and beyond}.
\newblock \bibinfo{journal}{\emph{Foundations and Trends{\textregistered} in Information Retrieval}} \bibinfo{volume}{3}, \bibinfo{number}{4} (\bibinfo{year}{2009}), \bibinfo{pages}{333--389}.
\newblock


\bibitem[Shao et~al\mbox{.}(2020)]%
        {bertpli20ijcai}
\bibfield{author}{\bibinfo{person}{Yunqiu Shao}, \bibinfo{person}{Jiaxin Mao}, \bibinfo{person}{Yiqun Liu}, \bibinfo{person}{Weizhi Ma}, \bibinfo{person}{Ken Satoh}, \bibinfo{person}{Min Zhang}, {and} \bibinfo{person}{Shaoping Ma}.} \bibinfo{year}{2020}\natexlab{}.
\newblock \showarticletitle{{BERT-PLI:} Modeling Paragraph-Level Interactions for Legal Case Retrieval}. In \bibinfo{booktitle}{\emph{Proceedings of the Twenty-Ninth International Joint Conference on Artificial Intelligence, {IJCAI} 2020}}, \bibfield{editor}{\bibinfo{person}{Christian Bessiere}} (Ed.). \bibinfo{publisher}{ijcai.org}, \bibinfo{pages}{3501--3507}.
\newblock
\urldef\tempurl%
\url{https://doi.org/10.24963/IJCAI.2020/484}
\showDOI{\tempurl}


\bibitem[Shi and Malik(2000)]%
        {spectral}
\bibfield{author}{\bibinfo{person}{Jianbo Shi} {and} \bibinfo{person}{Jitendra Malik}.} \bibinfo{year}{2000}\natexlab{}.
\newblock \showarticletitle{Normalized cuts and image segmentation}.
\newblock \bibinfo{journal}{\emph{IEEE Transactions on pattern analysis and machine intelligence}} \bibinfo{volume}{22}, \bibinfo{number}{8} (\bibinfo{year}{2000}), \bibinfo{pages}{888--905}.
\newblock


\bibitem[Singh et~al\mbox{.}(2023)]%
        {scieval23emnlp}
\bibfield{author}{\bibinfo{person}{Amanpreet Singh}, \bibinfo{person}{Mike D’Arcy}, \bibinfo{person}{Arman Cohan}, \bibinfo{person}{Doug Downey}, {and} \bibinfo{person}{Sergey Feldman}.} \bibinfo{year}{2023}\natexlab{}.
\newblock \showarticletitle{SciRepEval: A Multi-Format Benchmark for Scientific Document Representations}. In \bibinfo{booktitle}{\emph{Proceedings of the 2023 Conference on Empirical Methods in Natural Language Processing}}. \bibinfo{pages}{5548--5566}.
\newblock


\bibitem[Srivastava et~al\mbox{.}(2014)]%
        {dropout}
\bibfield{author}{\bibinfo{person}{Nitish Srivastava}, \bibinfo{person}{Geoffrey Hinton}, \bibinfo{person}{Alex Krizhevsky}, \bibinfo{person}{Ilya Sutskever}, {and} \bibinfo{person}{Ruslan Salakhutdinov}.} \bibinfo{year}{2014}\natexlab{}.
\newblock \showarticletitle{Dropout: a simple way to prevent neural networks from overfitting}.
\newblock \bibinfo{journal}{\emph{The journal of machine learning research}} \bibinfo{volume}{15}, \bibinfo{number}{1} (\bibinfo{year}{2014}), \bibinfo{pages}{1929--1958}.
\newblock


\bibitem[Sun et~al\mbox{.}(2023)]%
        {rankgpt23emnlp}
\bibfield{author}{\bibinfo{person}{Weiwei Sun}, \bibinfo{person}{Lingyong Yan}, \bibinfo{person}{Xinyu Ma}, \bibinfo{person}{Shuaiqiang Wang}, \bibinfo{person}{Pengjie Ren}, \bibinfo{person}{Zhumin Chen}, \bibinfo{person}{Dawei Yin}, {and} \bibinfo{person}{Zhaochun Ren}.} \bibinfo{year}{2023}\natexlab{}.
\newblock \showarticletitle{Is ChatGPT Good at Search? Investigating Large Language Models as Re-Ranking Agents}. In \bibinfo{booktitle}{\emph{Proceedings of the 2023 Conference on Empirical Methods in Natural Language Processing}}. \bibinfo{pages}{14918--14937}.
\newblock


\bibitem[Tsitsulin et~al\mbox{.}(2023)]%
        {dmon23jmlr}
\bibfield{author}{\bibinfo{person}{Anton Tsitsulin}, \bibinfo{person}{John Palowitch}, \bibinfo{person}{Bryan Perozzi}, {and} \bibinfo{person}{Emmanuel M{\"u}ller}.} \bibinfo{year}{2023}\natexlab{}.
\newblock \showarticletitle{Graph clustering with graph neural networks}.
\newblock \bibinfo{journal}{\emph{Journal of Machine Learning Research}} \bibinfo{volume}{24}, \bibinfo{number}{127} (\bibinfo{year}{2023}), \bibinfo{pages}{1--21}.
\newblock


\bibitem[Tworkowski et~al\mbox{.}(2024)]%
        {fot24nips}
\bibfield{author}{\bibinfo{person}{Szymon Tworkowski}, \bibinfo{person}{Konrad Staniszewski}, \bibinfo{person}{Miko{\l}aj Pacek}, \bibinfo{person}{Yuhuai Wu}, \bibinfo{person}{Henryk Michalewski}, {and} \bibinfo{person}{Piotr Mi{\l}o{\'s}}.} \bibinfo{year}{2024}\natexlab{}.
\newblock \showarticletitle{Focused transformer: Contrastive training for context scaling}.
\newblock \bibinfo{journal}{\emph{Advances in Neural Information Processing Systems}}  \bibinfo{volume}{36} (\bibinfo{year}{2024}).
\newblock


\bibitem[Wan and Yang(2008)]%
        {multidoc_cluslink08sigir}
\bibfield{author}{\bibinfo{person}{Xiaojun Wan} {and} \bibinfo{person}{Jianwu Yang}.} \bibinfo{year}{2008}\natexlab{}.
\newblock \showarticletitle{Multi-document summarization using cluster-based link analysis}. In \bibinfo{booktitle}{\emph{Proceedings of the 31st annual international ACM SIGIR conference on Research and development in information retrieval}}. \bibinfo{pages}{299--306}.
\newblock


\bibitem[Wu et~al\mbox{.}(2020)]%
        {pcfg20www}
\bibfield{author}{\bibinfo{person}{Zhijing Wu}, \bibinfo{person}{Jiaxin Mao}, \bibinfo{person}{Yiqun Liu}, \bibinfo{person}{Jingtao Zhan}, \bibinfo{person}{Yukun Zheng}, \bibinfo{person}{Min Zhang}, {and} \bibinfo{person}{Shaoping Ma}.} \bibinfo{year}{2020}\natexlab{}.
\newblock \showarticletitle{Leveraging passage-level cumulative gain for document ranking}. In \bibinfo{booktitle}{\emph{Proceedings of The Web Conference 2020}}. \bibinfo{pages}{2421--2431}.
\newblock


\bibitem[Wu et~al\mbox{.}(2019)]%
        {psgrole19sigir}
\bibfield{author}{\bibinfo{person}{Zhijing Wu}, \bibinfo{person}{Jiaxin Mao}, \bibinfo{person}{Yiqun Liu}, \bibinfo{person}{Min Zhang}, {and} \bibinfo{person}{Shaoping Ma}.} \bibinfo{year}{2019}\natexlab{}.
\newblock \showarticletitle{Investigating Passage-level Relevance and Its Role in Document-level Relevance Judgment}. In \bibinfo{booktitle}{\emph{Proceedings of the 42nd International {ACM} {SIGIR} Conference on Research and Development in Information Retrieval, {SIGIR} 2019, Paris, France, July 21-25, 2019}}, \bibfield{editor}{\bibinfo{person}{Benjamin Piwowarski}, \bibinfo{person}{Max Chevalier}, \bibinfo{person}{{\'{E}}ric Gaussier}, \bibinfo{person}{Yoelle Maarek}, \bibinfo{person}{Jian{-}Yun Nie}, {and} \bibinfo{person}{Falk Scholer}} (Eds.). \bibinfo{publisher}{{ACM}}, \bibinfo{pages}{605--614}.
\newblock
\urldef\tempurl%
\url{https://doi.org/10.1145/3331184.3331233}
\showDOI{\tempurl}


\bibitem[Xia and Wang(2023)]%
        {s2s23kdd}
\bibfield{author}{\bibinfo{person}{Jinxiong Xia} {and} \bibinfo{person}{Houfeng Wang}.} \bibinfo{year}{2023}\natexlab{}.
\newblock \showarticletitle{A Sequence-to-Sequence Approach with Mixed Pointers to Topic Segmentation and Segment Labeling}. In \bibinfo{booktitle}{\emph{Proceedings of the 29th ACM SIGKDD Conference on Knowledge Discovery and Data Mining}}. \bibinfo{pages}{2683--2693}.
\newblock


\bibitem[Xiao et~al\mbox{.}(2021)]%
        {lawfm21}
\bibfield{author}{\bibinfo{person}{Chaojun Xiao}, \bibinfo{person}{Xueyu Hu}, \bibinfo{person}{Zhiyuan Liu}, \bibinfo{person}{Cunchao Tu}, {and} \bibinfo{person}{Maosong Sun}.} \bibinfo{year}{2021}\natexlab{}.
\newblock \showarticletitle{Lawformer: A pre-trained language model for chinese legal long documents}.
\newblock \bibinfo{journal}{\emph{AI Open}}  \bibinfo{volume}{2} (\bibinfo{year}{2021}), \bibinfo{pages}{79--84}.
\newblock


\bibitem[Yang et~al\mbox{.}(2023)]%
        {baichuan23}
\bibfield{author}{\bibinfo{person}{Aiyuan Yang}, \bibinfo{person}{Bin Xiao}, \bibinfo{person}{Bingning Wang}, \bibinfo{person}{Borong Zhang}, \bibinfo{person}{Ce Bian}, \bibinfo{person}{Chao Yin}, \bibinfo{person}{Chenxu Lv}, \bibinfo{person}{Da Pan}, \bibinfo{person}{Dian Wang}, \bibinfo{person}{Dong Yan}, {et~al\mbox{.}}} \bibinfo{year}{2023}\natexlab{}.
\newblock \showarticletitle{Baichuan 2: Open large-scale language models}.
\newblock \bibinfo{journal}{\emph{arXiv preprint arXiv:2309.10305}} (\bibinfo{year}{2023}).
\newblock


\bibitem[Yang et~al\mbox{.}(2020)]%
        {smith20cikm}
\bibfield{author}{\bibinfo{person}{Liu Yang}, \bibinfo{person}{Mingyang Zhang}, \bibinfo{person}{Cheng Li}, \bibinfo{person}{Michael Bendersky}, {and} \bibinfo{person}{Marc Najork}.} \bibinfo{year}{2020}\natexlab{}.
\newblock \showarticletitle{Beyond 512 tokens: Siamese multi-depth transformer-based hierarchical encoder for long-form document matching}. In \bibinfo{booktitle}{\emph{Proceedings of the 29th ACM International Conference on Information \& Knowledge Management}}. \bibinfo{pages}{1725--1734}.
\newblock


\bibitem[Yang et~al\mbox{.}(2019)]%
        {xlnet19nips}
\bibfield{author}{\bibinfo{person}{Zhilin Yang}, \bibinfo{person}{Zihang Dai}, \bibinfo{person}{Yiming Yang}, \bibinfo{person}{Jaime Carbonell}, \bibinfo{person}{Russ~R Salakhutdinov}, {and} \bibinfo{person}{Quoc~V Le}.} \bibinfo{year}{2019}\natexlab{}.
\newblock \showarticletitle{Xlnet: Generalized autoregressive pretraining for language understanding}.
\newblock \bibinfo{journal}{\emph{Advances in neural information processing systems}}  \bibinfo{volume}{32} (\bibinfo{year}{2019}).
\newblock


\bibitem[Yu et~al\mbox{.}(2023)]%
        {topiccohere_23emnlp}
\bibfield{author}{\bibinfo{person}{Hai Yu}, \bibinfo{person}{Chong Deng}, \bibinfo{person}{Qinglin Zhang}, \bibinfo{person}{Jiaqing Liu}, \bibinfo{person}{Qian Chen}, {and} \bibinfo{person}{Wen Wang}.} \bibinfo{year}{2023}\natexlab{}.
\newblock \showarticletitle{Improving Long Document Topic Segmentation Models With Enhanced Coherence Modeling}. In \bibinfo{booktitle}{\emph{Proceedings of the 2023 Conference on Empirical Methods in Natural Language Processing}}. \bibinfo{pages}{5592--5605}.
\newblock


\bibitem[Yu et~al\mbox{.}(2022)]%
        {optmatch22coling}
\bibfield{author}{\bibinfo{person}{Weijie Yu}, \bibinfo{person}{Liang Pang}, \bibinfo{person}{Jun Xu}, \bibinfo{person}{Bing Su}, \bibinfo{person}{Zhenhua Dong}, {and} \bibinfo{person}{Ji{-}Rong Wen}.} \bibinfo{year}{2022}\natexlab{}.
\newblock \showarticletitle{Optimal Partial Transport Based Sentence Selection for Long-form Document Matching}. In \bibinfo{booktitle}{\emph{Proceedings of the 29th International Conference on Computational Linguistics, {COLING} 2022, Gyeongju, Republic of Korea, October 12-17, 2022}}, \bibfield{editor}{\bibinfo{person}{Nicoletta Calzolari}, \bibinfo{person}{Chu{-}Ren Huang}, \bibinfo{person}{Hansaem Kim}, \bibinfo{person}{James Pustejovsky}, \bibinfo{person}{Leo Wanner}, \bibinfo{person}{Key{-}Sun Choi}, \bibinfo{person}{Pum{-}Mo Ryu}, \bibinfo{person}{Hsin{-}Hsi Chen}, \bibinfo{person}{Lucia Donatelli}, \bibinfo{person}{Heng Ji}, \bibinfo{person}{Sadao Kurohashi}, \bibinfo{person}{Patrizia Paggio}, \bibinfo{person}{Nianwen Xue}, \bibinfo{person}{Seokhwan Kim}, \bibinfo{person}{Younggyun Hahm}, \bibinfo{person}{Zhong He}, \bibinfo{person}{Tony~Kyungil Lee}, \bibinfo{person}{Enrico Santus}, \bibinfo{person}{Francis Bond}, {and} \bibinfo{person}{Seung{-}Hoon Na}} (Eds.). \bibinfo{publisher}{International
  Committee on Computational Linguistics}, \bibinfo{pages}{2363--2373}.
\newblock
\urldef\tempurl%
\url{https://aclanthology.org/2022.coling-1.208}
\showURL{%
\tempurl}


\bibitem[Zhai(2008)]%
        {statIR}
\bibfield{author}{\bibinfo{person}{ChengXiang Zhai}.} \bibinfo{year}{2008}\natexlab{}.
\newblock \showarticletitle{Statistical language models for information retrieval}.
\newblock \bibinfo{journal}{\emph{Synthesis lectures on human language technologies}} \bibinfo{volume}{1}, \bibinfo{number}{1} (\bibinfo{year}{2008}), \bibinfo{pages}{1--141}.
\newblock


\bibitem[Zhai et~al\mbox{.}(2015)]%
        {subtopic_retrieval_15}
\bibfield{author}{\bibinfo{person}{ChengXiang Zhai}, \bibinfo{person}{William~W Cohen}, {and} \bibinfo{person}{John Lafferty}.} \bibinfo{year}{2015}\natexlab{}.
\newblock \showarticletitle{Beyond independent relevance: methods and evaluation metrics for subtopic retrieval}. In \bibinfo{booktitle}{\emph{Acm sigir forum}}, Vol.~\bibinfo{volume}{49}. ACM New York, NY, USA, \bibinfo{pages}{2--9}.
\newblock


\bibitem[Zheng et~al\mbox{.}(2019)]%
        {subtopic_mdsum_19_emnlp}
\bibfield{author}{\bibinfo{person}{Xin Zheng}, \bibinfo{person}{Aixin Sun}, \bibinfo{person}{Jing Li}, {and} \bibinfo{person}{Karthik Muthuswamy}.} \bibinfo{year}{2019}\natexlab{}.
\newblock \showarticletitle{Subtopic-driven multi-document summarization}. In \bibinfo{booktitle}{\emph{Proceedings of the 2019 conference on empirical methods in natural language processing and the 9th international joint conference on natural language processing (EMNLP-IJCNLP)}}. \bibinfo{pages}{3153--3162}.
\newblock


\bibitem[Zhou et~al\mbox{.}(2020)]%
        {cda20emnlp}
\bibfield{author}{\bibinfo{person}{Xuhui Zhou}, \bibinfo{person}{Nikolaos Pappas}, {and} \bibinfo{person}{Noah~A. Smith}.} \bibinfo{year}{2020}\natexlab{}.
\newblock \showarticletitle{Multilevel Text Alignment with Cross-Document Attention}. In \bibinfo{booktitle}{\emph{Proceedings of the 2020 Conference on Empirical Methods in Natural Language Processing, {EMNLP} 2020, Online, November 16-20, 2020}}, \bibfield{editor}{\bibinfo{person}{Bonnie Webber}, \bibinfo{person}{Trevor Cohn}, \bibinfo{person}{Yulan He}, {and} \bibinfo{person}{Yang Liu}} (Eds.). \bibinfo{publisher}{Association for Computational Linguistics}, \bibinfo{pages}{5012--5025}.
\newblock
\urldef\tempurl%
\url{https://doi.org/10.18653/V1/2020.EMNLP-MAIN.407}
\showDOI{\tempurl}


\bibitem[Zhou et~al\mbox{.}(2023)]%
        {legalsum23sigir_ap}
\bibfield{author}{\bibinfo{person}{Youchao Zhou}, \bibinfo{person}{Heyan Huang}, {and} \bibinfo{person}{Zhijing Wu}.} \bibinfo{year}{2023}\natexlab{}.
\newblock \showarticletitle{Boosting legal case retrieval by query content selection with large language models}. In \bibinfo{booktitle}{\emph{Annual International {ACM} {SIGIR} Conference on Research and Development in Information Retrieval in the Asia Pacific Region, {SIGIR-AP} 2023, Beijing, China, November 26-28, 2023}}, \bibfield{editor}{\bibinfo{person}{Qingyao Ai}, \bibinfo{person}{Yiqin Liu}, \bibinfo{person}{Alistair Moffat}, \bibinfo{person}{Xuanjing Huang}, \bibinfo{person}{Tetsuya Sakai}, {and} \bibinfo{person}{Justin Zobel}} (Eds.). \bibinfo{publisher}{{ACM}}, \bibinfo{pages}{176--184}.
\newblock
\urldef\tempurl%
\url{https://doi.org/10.1145/3624918.3625328}
\showDOI{\tempurl}


\bibitem[Zhuang et~al\mbox{.}(2023)]%
        {rankt5_23sigir}
\bibfield{author}{\bibinfo{person}{Honglei Zhuang}, \bibinfo{person}{Zhen Qin}, \bibinfo{person}{Rolf Jagerman}, \bibinfo{person}{Kai Hui}, \bibinfo{person}{Ji Ma}, \bibinfo{person}{Jing Lu}, \bibinfo{person}{Jianmo Ni}, \bibinfo{person}{Xuanhui Wang}, {and} \bibinfo{person}{Michael Bendersky}.} \bibinfo{year}{2023}\natexlab{}.
\newblock \showarticletitle{Rankt5: Fine-tuning t5 for text ranking with ranking losses}. In \bibinfo{booktitle}{\emph{Proceedings of the 46th International ACM SIGIR Conference on Research and Development in Information Retrieval}}. \bibinfo{pages}{2308--2313}.
\newblock


\end{thebibliography}
